\documentclass[aps,prd,amsmath,amssymb,11pt]{revtex4}
\usepackage{graphicx}% Include figure files
\usepackage{bm}% bold math
\def\journal#1#2#3#4{{#1} {\bf #2}, #3 (#4)}
\newcommand{\be}{\begin{equation}}
\newcommand{\ee}{\end{equation}}
\newcommand{\bea}{\begin{eqnarray}}
\newcommand{\eea}{\end{eqnarray}}
\newcommand{\hf}{\frac12}
\newcommand{\nn}{\nonumber\\}
\def\eq#1{(\ref{#1})}
\def\la{\langle}
\def\ra{\rangle}
\def\tr{{\mathrm{tr}}}
\def\Tr{{\mathrm{Tr}}}
\def\ord#1{{\cal O}\left(#1\right)}
\def\mr#1{{\mathrm{#1}}}
\def\v#1{{\bm{#1}}}
\def\fd#1#2{\frac{\delta#1}{\delta#2}}
\def\fdd#1#2#3{\frac{\delta^2#1}{\delta#2\delta#3}}
\def\dt{{\Delta t}}
\def\hj{{\hat j}}
\def\hphi{{\hat\phi}}
\def\hpsi{{\hat\psi}}

\def\hs{\hat\sigma}
\def\hx{\hat x}
\def\hy{\hat y}
\def\hD{{\hat D}}

\def\ih{\frac{i}{\hbar}}

\def\sign{\mr{sign}}

\begin{document}
\title{Classical and quantum effective theories}
\author{Janos Polonyi}
\email{polonyi@iphc.cnrs.fr}
\affiliation{Strasbourg University, CNRS-IPHC, 23 rue du Loess, BP28 67037 Strasbourg Cedex 2, France}

\begin{abstract}
A generalization of the action principle of classical mechanics, motivated by the closed time path scheme of quantum field theory, is presented to deal with initial condition problems and dissipative forces. The similarities of the classical and the quantum cases are underlined. In particular, effective interactions that describe classical dissipative forces represent the system-environment entanglement. The relation between the traditional effective theories and their CTP extension is briefly discussed, and a few qualitative examples are mentioned.
\end{abstract}
\maketitle
%\tableofcontents

\section{Introduction}
An elementary theory is supposed to describe fundamental degrees of freedom. It is always assumed that the dynamics is closed, the classical equations of motions can be derived from an action and the quantum states are represented by vectors in a linear space with unitary time evolution. In constructing an effective theory for some observed dynamical variables, called the system, we admit that their dynamics is open and their interactions with the unobserved rest of the Universe, the environment, render the effective interactions highly involved. In fact, effective theories are supposed to encompass dissipative forces and mixed states. Some of the analytical tools that have been developed for elementary theories, namely the classical action principle and the use of transition amplitudes between pure states in the quantum case, are therefore insufficient for effective theories.

Effective theories are discussed first on the classical level in this paper. A generalization of the elementary, holonomic forces acting in closed systems, called semiholonomic forces, is given. The extension is, on the one hand, restricted enough still to find an action principle to deal with them and, on the other hand, sufficiently general to cover effective forces that can be induced by holonomic system-environment forces. It turns out that these effective forces can always be given account of by creating a copy of the system and letting it interact with the original system with holonomic forces. The method to deal with effective, open quantum systems is motivated by the extension of the naive quantization rules and the density matrix as a general framework to represent probability. The result is the already well known closed time path (CTP) formalism of quantum theory.

The possibility of mapping the system-environment interactions into the interactions of two copies of the system is implicitly present in quantum theory, in the perturbation expansion for the Heisenberg representation \cite{schw}. The ensuing CTP formalism has already been used extensively in condensed matter physics \cite{keldysh,smith,chou,kamenev} and quantum field theory \cite{umezawa,vilkovisky,calzetta}, and has recently been mentioned in the context of classical mechanics \cite{arrow,galley}. The CTP formalism is presented here as a natural, simple extension of the classical action principle to deal with effective theories in both classical and quantum mechanics. 

The discussion of the quantum case is to underline the uniformity of the formalism and contains a few new details only: A simple perturbative proof is given that  CTP effective theories extend the traditional effective field theories \cite{weinberg,leutwyler} to cover processes that leave the environment in an excited state, and this extension is demonstrated by mentioning inclusive scattering probabilities. Furthermore, it is shown that causality is not satisfied automatically but follows if the time evolution is unitarity in a discrete spectrum. Finally a few general remarks are mentioned about the decoherence, entanglement, classical and quantum fluctuations, and the classical limit. 

The discussion of classical effective theories in Sec. \ref{clmechs} starts with a brief description of two new features an effective theory should possess. It should handle initial, rather than boundary, condition problems because we do not know the final state of an unobserved environment. This is the subject of Sec. \ref{intics}. Furthermore, it should handle semiholonomic forces, which arise from holonomic system-environment interactions. The extension of the action principle by reduplicating the degrees of freedom to cover such forces is presented in Sec. \ref{semihs}. A further extension of this latter action principle for initial condition problems, the classical CTP formalism, is presented in Sec. \ref{initcs} by introducing a symplectic structure for the two CTP copies and by joining their trajectories at the final time. This scheme is the constructive proof that the class of semiholonomic forces is closed with respect to generating effective interactions. One of the main advantages of the CTP action principle, the straightforward introduction of the retarded Green function, is demonstrated in Sec. \ref{cgreens}. Section \ref{ceffths} contains the discussion of the effective action. The distinguishing piece of the effective action that describes the coupling of the two CTP copies of the system, the influence functional, is shown to be related to semiholonomic forces in Sec. \ref{ccctpcs}.

The second part of this work, Sec. \ref{qctpds}, is devoted to quantum dynamics. The simple quantization rules, the path integral representation of expectation values rather than transition amplitudes, and the Green functions of open systems are mentioned briefly in Sec. \ref{semiholq}. The quantum effective theories are the subject of Sec. \ref{qeffths} where the effective action is given as a sum of one and two CTP copies contributions. The former agrees with the effective action of traditional quantum field theories where the initial and final states are the vacuum. The latter, the influence functional, arises from processes that leave the environment in an excited state at the final time. Two examples of using CTP effective theories follow after that. It is shown in Sec. \ref{scatts} that inclusive scattering processes, where certain final particles remain unobserved, are more natural to be described by an effective CTP theory. The causal structure of interactions is not a trivial result in effective theories, and the realization of causality and its possible violation are discussed in Sec. \ref{causals}. The possibility to calculate the reduced density matrix by means of path integration opens the way to address some outstanding problems of the quantum-classical crossover. A few qualitative remarks about decoherence, entanglement, the relation of classical and quantum fluctuations and the classical limit are made in Sec. \ref{qctrs}. Finally, Sec. \ref{sums} contains the summary. 

Two Appendixes are added to make this work more self-contained. The way semiholonomic forces violate the usual conservation law is discussed in Appendix \ref{nothera} by checking the status of Noether's theorem in effective theories in some simple cases. The system-environment entanglement leads to a mixed system state. This can be followed in the clearest manner by using relative states, defined in Appendix \ref{relsta}.

\section{Classical mechanics}\label{clmechs}
The classical effective dynamics will be discussed within a generic model where $N_s$ system and $N_e$ environment coordinates, $x$  and $y$, respectively, obey a closed dynamics that is described by a Lagrangian, $L(x,y,\dot x,\dot y)$. The trajectories for the time interval $t_i<t<t_f$ are the solutions of the equations of motion, $\ddot x=F_s(x,\dot x,y,\dot y)$, and $\ddot y=F_e(x,\dot x,y,\dot y)$, 
subject of some auxiliary conditions, $c_s(x(t_s),\dot x(t_s))=0$, $c_e(y(t_e),\dot y(t_e))=0$, where $t_s,t_e=t_i$ or $t_f$. Since the environment is not observed it is natural to use $t_e=t_f$. The effective system equation of motion, $\ddot x=F_s(x,\dot x,y[x;c_e],\dot y[x;c_e])$, can simply be found by inserting the solution $y=y[x;c_e]$ of the environment equation for a general system trajectory into the system equation. Our goal is to find the effective action that generates the effective system equation as a variational equation.

\subsection{Initial conditions in effective theories}\label{intics}
The first problem is related to auxiliary conditions. The general solution of the equations of motion contains $2(N_s+N_e)$ free parameters: hence the effective equation must be an integrodifferential equation of order $2(N_s+N_e)$. But we cannot solve this equation if we possess $2N_s$ auxiliary conditions only. One can recover the manifold of all solutions by exploiting the $2(N_s+N_e)$ free parameters of the solution of the effective equation but an effective theory should always be equipped with a prescription or some additional information to select the $2N_s$-dimensional submanifold to cover the physically realizable system trajectories.

The attractive feature of the variation principle is its efficiency to find constrained extrema by Legendre transformation. This scheme offers a solution to our problem, namely the system trajectory can formally be defined by functional derivation rather than solving a higher order differential equation. For this end we introduce a formal source, $j(t)$, to diagnose the effective dynamics by using the action $S[x,y;j]=S[x,y]+jx$, where $S[x,y]$ denotes the action of the full, closed system and the scalar product of time dependent functions, $f(t)$, $g(t)$, is $fg=\int dtf(t)g(t)$. Furthermore, we define the functional $W[j]=S[x,y,;j]$, where the trajectories $x$ and $y$ are eliminated by solving the variational equation of motion of $W$, considered as an action functional of the trajectories for fixed $j$, 
\be
\fd{S}{x}+j=0,~~~\fd{S}{y}=0,
\ee
subject of the initial conditions, $c_s(x(t_i),\dot x(t_i))=0$, $c_e(y(t_i),\dot y(t_i))=0$.
The functional derivative
\be\label{fltrvtr}
\fd{W[j]}{j}=\fd{S}{x}\fd{x}{j}+\fd{S}{y}\fd{y}{j}+\fd{x}{j}j+x=x
\ee
shows that the knowledge of $W[j]$ is sufficient to find the desired trajectory.

This construction gives more than an algorithm to find the system trajectory: it provides us with the effective action, $S_{eff}[x]=W[j]-xj$, where $x$ is defined by Eq. \eq{fltrvtr}. In fact, its variational equation,
\be
\fd{S_{eff}}{x}=\fd{W}{j}\fd{j}{x}-x\fd{j}{x}-j=-j,
\ee
is satisfied by the system trajectory for vanishing $j$. This way of obtaining the effective action yields
\be\label{neffa}
S_{eff}[x]=S[x,y[x,c_e]]
\ee
by construction. Another advantage of this construction is that it is available even if the effective theory is needed for such a combination of the coordinates that is not present in the original action. 

The argument presented above is formal and hides a problem, related to auxiliary conditions: Though one usually possesses only the initial conditions for the unobserved environment the action principle cannot handle initial value problems. In fact, let us discretize the time by replacing the trajectory, $x(t)$, by a set of numbers, $x_j=x(t_i+j\dt)$, $j=1,\ldots,n=(t_f-t_i)/\dt$ and use $\partial S/\partial x_k=0$, $0<k<n$ as the equation of motion. If we set the value of the initial velocity, $v_i=(x_1-x_0)/\dt$ beside the initial coordinate, $x_i=x_0$, then we have to impose the equation of motion at the final time to have a sufficient number of equations to determine the trajectory. But it is well known that the variation of the action with respect to the final point of the trajectory that solves the equation of motion is the generalized momentum and its vanishing is an unacceptable constraint. A modification of the action principle is needed where the equation of motion at the final time is trivial, $0=0$.

\subsection{Semiholonomic forces}\label{semihs}
Another problem to solve in effective theories is that the observed system is open and the Lagrangian description does not apply. The traditional description of a closed system starts with the d'Alembert principle, stating that the virtual work during a variation, $\delta x$, of the external force, $F$, acting on a particle plus the inertial force, $-m\ddot x$, is vanishing, 
\be
(F-m\ddot x)\delta x=0.
\ee
The main assumption, which renders the methods of analytical mechanics powerful, is that the external force, which depends on the coordinate and the velocity, is holonomic: i.e., the virtual work of the external force during the variation $x(t)\to x(t)+\delta x(t)$ can be written at a given time in terms of the derivatives of a scalar potential, $U(x,\dot x)$,
\be\label{dalf}
F(x,\dot x)\delta x=-\delta x\partial_xU(x,\dot x)-\delta\dot x\partial_{\dot x}U(x,\dot x).
\ee
This equation involves not only the variation of the coordinate, $\delta x(t)$, but the variation of the velocity, $\delta\dot x(t)$, too. This is an important point, as it forces us to view the dynamical problem globally, in terms of the trajectory $x(t)$, rather than locally, at a given time. The relation 
\be\label{varid}
\delta\dot x=\frac{d}{dt}\delta x
\ee
follows in this manner. This identity can be build into our equations if, by following Hamilton, we integrate d'Alembert's principle in time,
\be
0=-\int_{t_i}^{t_f}dt[\delta x(m\ddot x+\partial_xU)+\delta\dot x\partial_{\dot x}U].
\ee
This equation  can be written by the use of the identity \eq{varid} as a variational equation,
\be\label{hamve}
0=\delta\int_{t_i}^{t_f}dtL-\delta x(m\dot x+\partial_{\dot x}U)\biggr|_{t_i}^{t_f},
\ee
involving the Lagrangian $L(x,\dot x)=m\dot x^2/2-U(x,\dot x)$. Note in passing that the last term in \eq{hamve} cancels the boundary contribution arising from the calculation of the first term, contrary to a widespread presentation of the Lagrangian formalism.

We generalize this procedure for semiholonomic forces, defined by Eq. \eq{dalf} with the modification that the derivatives act on a subset of the coordinates only. For this end we use two copies of the coordinate, $x\to\hx=(x^+,x^-)$, where $x^+$ is called the active coordinate and $x^-$, the passive coordinate represents the nonholonomic forces of the environment, and assume the form
\be
F(x,\dot x)\delta x=-[\delta x\partial_{x^+}U(\hx,\dot{\hx})+\delta\dot x\partial_{\dot x^+}U(\hx,\dot{\hx})]_{|x^+=x^-=x}
\ee
for the virtual work. Since $x^-(t)$ describes the same motion as $x^+(t)$ it is reasonable to homogenize the formalism by performing independent variation on $x^+$ and $x^-$. This gives the idea to extend the redoubling of the coordinates in the potential energy to the kinetic energy, as well, and to rearrange things in such a manner that we arrive at two equivalent variational equations. The on-shell condition,
\be\label{onshell}
x^+(t)=x^-(t),
\ee
is automatically satisfied in this manner. 

This plan can be realized by using a generalized Lagrangian, $L(\hx,\dot{\hx})$, and by requiring that the equation of motion be identical for $x^+$ and $x^-$. A sufficient condition to meet this condition is to have a Lagrangian that is multiplied by a sign only when the active and passive coordinates are exchanged,
\be\label{ctps}
L(\tau\hx,\tau\dot{\hx})=\pm L(\hx,\dot{\hx}),
\ee
where $\tau(x^+,x^-)=(x^-,x^+)$.

The symmetry under $\hx\to\tau\hx$ indicates a redundancy of the formalism, and it is reminiscent of a gauge transformation. The extension with the sign $+$ in Eq. \eq{ctps} gives nothing new, but it corresponds to a system where every degree of freedom exists in two copies and they interact with holonomic forces. The choice if the sign $-$ introduces a new, nontrivial symplectic structure, to be exploited below.

\subsection{CTP action principle}\label{initcs}
We have seen so far two problems that were posed by the initial conditions and the nonholonomic forces. It will now be shown that the extended action principle, outlined above, solves both of them if the sign $-$ is used in Eq. \eq{ctps}. A side product of the argument will be the proof that effective forces, generated by holonomic forces are semiholonomic forces.

We start with a simpler problem, the application of the extended Lagrangian formalism of Sec. \ref{semihs} for the initial conditions in the case of holonomic forces only. The Lagrangian, $L(x,\dot x)$, gives rise to $L(\hx,\dot{\hx})=L(x^+,\dot x^+)\pm L(x^-,\dot x^-)$, and by imposing the same initial conditions we satisfy Eq. \eq{onshell}. How do we render the equation of motion trivial at the final time without losing the well-defined nature of the trajectories? We might renounce the equation of motion for $x^-$ altogether if we would impose Eq. \eq{onshell}, the result of the symmetry \eq{ctps}. But it is better to keep the treatment of the two copies of the coordinates on equal footing and retain the redundant equations of motion except at the final time where Eq. \eq{onshell} is invoked,
\be\label{ctpfc}
x^+(t_f)=x^-(t_f).
\ee
The variational principle is now defined within the space of trajectory pairs that satisfy identical initial condition, $c(x^\pm,\dot x^\pm)=0$, and the final condition \eq{ctpfc}. The variation of the action, evaluated on the solution of the equations of motion, with respect to the final coordinate within this space of trajectories is vanishing by construction if the sign $-$ is chosen in Eq. \eq{ctps}. Therefore we define the action as \cite{arrow,galley}
\be\label{cctpacte}
S[\hx]=S[x^+]-S[x^-],
\ee
where $S[x]$ denotes the traditional action of the Lagrangian $L(x,\dot x)$.

This reduplication of the degrees of freedom and the resulting variational principle can be reached in another, equivalent manner, where we replay the motion backward in time. The original trajectory, $x(t)$, defined for $t_i<t<t_f$, and satisfying the initial condition $c(x,\dot x)=0$ is now extended to
\be\label{ttraj}
\tilde x(\tilde t)=\begin{cases}x(\tilde t)&t_i<\tilde t<t_f,\cr x(2t_f-\tilde t)&t_f<\tilde t<2t_f-t_i,\end{cases}
\ee
and the system returns to its initial conditions after having followed a closed path during time $2(t_f-t_i)$. The reduplication of the time is represented as a reduplication of the degrees of freedom,
\be\label{tildectp}
\begin{pmatrix}x^+(t)\cr x^-(t)\end{pmatrix}
=\begin{pmatrix}\tilde x(t)\cr\tilde x(2t_f-t)\end{pmatrix},
\ee
and the action is given by
\be
S[\hx]=\int_{t_i}^{t_f}dtL(x^+(t),\dot x^+(t))+\int_{t_f}^{t_i}dtL(x^-(t),\dot x^-(t)),
\ee
where the limit of the integration in the second term on the right hand side indicates that the motion is followed in that section backward in time time. The second time inversion, carried out implicitly on $x^-(t)$ in Eq. \eq{tildectp}, returns the same direction of time for $x^+(t)$ and $x^-(t)$. 

We want to bring a further modification of the action because it is degenerate for $x^+(t)=x^-(t)$. To arrive at well defined Green functions we split this degeneracy by redefining the action,
\be\label{ctpactspl}
S[\hx]=S_0[x^+]-S_0[x^-]+S_{spl}[\hx],
\ee
with a suitable chosen, infinitesimal $S_{spl}[\hx]$. The choice of an imaginary splitting term, for instance
\be\label{splitact}
S_{spl}[\hx]=i\frac\epsilon2\int_{t_i}^{t_f}dt[(x^+(t))^2+(x^-(t))^2],
\ee
with $\epsilon=0^+$, has the advantage that the condition \eq{onshell} is satisfied by $\Re\hx(t)$ after solving the equation of motion. The  exchange of the two CTP copies, the CTP conjugation, transforms the action as
\be\label{ctpsym}
S[\hx]=-S^*[\tau\hx],
\ee
and the symmetry of the equation of motion with respect to this transformation will be called CTP symmetry. Another advantage of the splitting term \eq{splitact}, to be verified below, is that the Green functions derived from this action correspond to the generic initial condition $x(t_i)=\dot x(t_i)=0$. Note that the actual choice of the final time, $t_f$, when the motion is turned backward in time is arbitrary, the CTP trajectory, $\hx(t)$ is independent of $t_f$ as long as $t_f>t$.

The extension of the CTP scheme to the Hamiltonian formalism is straightforward. The generalized momenta are defined by
\be
p^\pm=\fd{L(\hx,\dot{\hx})}{\dot x^\pm}=\pm\fd{L(x^\pm,\dot x^\pm)}{\dot x^\pm}
\ee
and the Hamiltonian is
\be\label{ctpham}
H(\hx,\hat p)=\dot{\hx}\hat p-L(\hx,\dot{\hx})=H(x^+,p^+)-H(x^-,p^-),
\ee
where $H(x,p)=\dot xp-L(x,\dot x)$.

The status of causality is not as trivial in the CTP formalism as in the case of Newton's equation. In fact, let us couple an external source, $j$, to the coordinate linearly by extending the Lagrangian, $L(x,\dot x)\to L(x,\dot x)+jx$. The solution of initial value problems is always causal in a finite system, and the influence of the source $j(t)=j_0\delta(t-t_0)$ is for $t>t_0$. In the CTP formalism we have $\hj=j_0\delta(t-t_0)(1,1)$ which modifies the trajectory $\tilde x(\tilde t)$ in the time interval $t_0<\tilde t<2t_f-t_i-t_0$ for a causal dynamics. This feature can easily be proven by recalling that one imposes time reverse auxiliary conditions at $\tilde t=t_i$, and $2t_f-t_i$. But this argument is available only if we possess all the information needed to render the equation of motion unique, a nontrivial problem in effective theories in the case of an infinite environment, $N_e\to\infty$, according to Sec. \ref{intics}. We return to this question in Sec. \ref{causals}.

Symmetries can be extended in a direct or a CTP conjugated manner. In fact, if one considers a symmetry transformation $x\to x'$ that preserves the action, then its CTP realization can be either $\hx\to\hx'$, $\Re S\to\Re S$ (direct) or $\hx\to\tau\hx'$, $\Re S\to-\Re S$ (CTP conjugated). If the time is changed, as well, then the transformation acts on the trajectories and the auxiliary conditions and the action may pick up a boundary term, as usual. In particular, the time reversal transformation, $t\to t^T=t_f+t_i-t$, $x(t)\to x^T(t)=x(t^T)$, $c^T(z,\dot z)=c(z^T,-\dot z^T)$, is chosen to be represented with CTP conjugation because the action can be preserved in this manner.

\subsection{Green functions}\label{cgreens}
The Green functions are defined with the help of an external source, $S[\hx]\to S[\hx;\hj]=S[\hx]+\hj\hs\hx$, where the metric tensor of the CTP symplectic structure, 
\be
\hs=\begin{pmatrix}1&0\cr0&-1\end{pmatrix},
\ee
is introduced. The Legendre transform of the action, $W[\hj]=S[\hx;\hj]$, with the independent variable $\hj$ is defined by substituting the solution of the variational equation of $S[\hx;\hj]$,
\be\label{heom}
\fd{S}{\hx}=-\hs\hj,
\ee
and the auxiliary conditions into $S[\hx;\hj]$. The Green functions, $\hD(t_1,\ldots,t_n)$, are read off from the functional Taylor expansion,
\be\label{ctpgfnctd}
W[\hj]=\sum_{n=0}^\infty\frac1{n!}\sum_{\sigma_1,\ldots,\sigma_n}\sigma_1\cdots\sigma_n\int dt_1\cdots dt_nD^{\sigma_1,\ldots,\sigma_n}(t_1,\ldots,t_n)j^{\sigma_1}(t_1)\cdots j^{\sigma_n}(t_n).
\ee
The identity
\be
\fd{W}{\hj}=\hs\hx
\ee
shows that the $n$th order Green function, $D^{\sigma_1,\ldots,\sigma_n}(t_1,\ldots,t_n)$, represents the $\ord{\hj^{n-1}}$ contributions of the trajectory. In particular, the only Green function of a harmonic system,
\be\label{cctphoa}
S_0[\hx]=\hf\hx\hat K_0\hx,
\ee
is a two-point function, $\hD_0=\hat K_0^{-1}$, and the solution of the equation of motion \eq{heom} with a CTP symmetric external source, $j^\pm(t)=j(t)$, is
\be\label{hamsol}
x(t)=-\sum_{\sigma'}\int dt'D_0^{\sigma\sigma'}(t,t')\sigma'j(t').
\ee
This equation holds for arbitrary $\sigma$, and therefore the relation
\be\label{cctpid}
D^{++}+D^{--}=D^{+-}+D^{-+}
\ee
follows for two-point functions.

The CTP symmetry of the physical trajectory imposes the block structure $\hat K=\hs{\cal C}_4[K^n,K^f,K^i_1,K^i_2]\hs$, where
\be
{\cal C}_4[K^n,K^f,K^i_1,K^i_2]=\begin{pmatrix}K^n+iK^i_1&-K^f+iK^i_2\cr K^f+iK^i_2&-K^n+iK^i_1\end{pmatrix},
\ee
containing four real functions, $K^n$, $K^f$, $K^i_1$, and $K^i_2$. We can safely assume that $\hat K$ is symmetric, thus $K^{i\tr}_j=K^i_j$, $K^{n\tr}=K^n$, and $K^{ftr}=-K^f$. The inversion of $\hat K$, together with the condition \eq{cctpid} yields $K^i_1=K^i_2$, and the form $\hD={\cal C}_3[D^n,D^f,D^i]$ with
\be
{\cal C}_3[D^n,D^f,D^i]=\begin{pmatrix}D^n+iD^i&-D^f+iD^i\cr D^f+iD^i&-D^n+iD^i\end{pmatrix},
\ee
is found for the CTP Green function in terms of three real functions, $D^{i\tr}=D^i$, $D^{n\tr}=D^n$, and $D^{ftr}=-D^f$. The combinations $K^{\stackrel{r}{a}}=K^n\pm K^f$ and $D^{\stackrel{r}{a}}=D^n\pm D^f$ establish the relation between $\hat K$ and $\hD$,
\bea\label{ctppropinv}
K^{\stackrel{r}{a}}&=&\left(D^{\stackrel{r}{a}}\right)^{-1},\nn
K^i&=&-(D^a)^{-1}D^i(D^r)^{-1}.
\eea

If the initial conditions $x(t_i)=\dot x(t_i)=0$ are used in the construction of the generator functional $W[\hj]$, then $D^r$ and $D^a$ are indeed the retarded and advanced Green functions, and hence $D^n$ and $D^f$ will be called near and far Green functions. Note that the imaginary part of the Green function, induced by the piece \eq{splitact} in the action, drops out from the classical trajectory \eq{hamsol}, and its detailed form is not important.

The Green function of a harmonic oscillator,
\be\label{holagr}
L=\frac{m}2\dot x^2-\frac{m\Omega^2}2x^2,
\ee
with the initial conditions $x(t_i)=\dot x(t_i)=0$,
\be
\hD_0(t,t')=\int\frac{d\omega}{2\pi}e^{-i\omega(t-t')}\hD_0(\omega),
\ee
is given by $\hD_0(\omega)={\cal C}_3[D^n_0(\omega),D^f_0(\omega),D^i_0(\omega)]$ with
\bea\label{hogfnct}
D_0^n(\omega)&=&\frac1mP\frac1{\omega^2-\Omega^2},\nn
D_0^f(\omega)&=&-\frac{i\pi}m\sign(\omega)\delta(\omega^2-\Omega^2),\nn
D_0^i(\omega)&=&-\frac\pi{m}\delta(\omega^2-\Omega^2),
\eea
in the limit $t_i\to-\infty$, $t_f\to\infty$, carried out to recover translation invariance in time \cite{effthpch}. The pole structure in frequency space assures causality, $D_0^r(t,t')=0$ for $t<t'$ and $D_0^a(t,t')=0$ for $t>t'$. The use of the representation $\delta_\epsilon(\omega)=\epsilon/\pi(\omega^2+\epsilon^2)$ of the Dirac delta makes the inversion trivial and yields 
\be\label{hoigfnct}
K_0^n=m(\omega^2-\Omega^2),~~~K_0^f=i\sign(\omega)\epsilon,~~~K_0^i=\epsilon.
\ee
The inverse of the Green function, $\hat K_0$, can be used in Eq. \eq{cctphoa}, giving
\be\label{acctpact}
S_0[\hx]=S_0[x^+]-S_0[x^-]+S_{BC}[\hx]
\ee
where $S_{BC}[\hx]=S_{spl}[\hx]+S_f[\hx]$ handles the boundary conditions. The initial conditions are fixed by $S_{spl}[\hx]$, which is given by Eq. \eq{splitact} with $t_i=-\infty$, $t_f=\infty$. The new part,
\bea\label{ctpfcf}
S_f[\hx]&=&-i\frac\epsilon\pi\int d\omega\Theta(\omega)x^{-*}(\omega)x^+(\omega)\nn
&=&\frac\epsilon\pi\int_{-\infty}^\infty dtdt'\frac{x^+(t)x^-(t')}{t-t'+i\epsilon}
\eea
represents the final condition \eq{ctpfc} by a nonlocal, time translation invariant, infinitesimal coupling between the CTP copies. The first equation, where the frequency integral is for $-\infty<\omega<\infty$, shows that the positive frequency components of $x^+$ are coupled to the negative frequency components of $x^-$ at $t_f=\infty$. In other words, the actual form of $S_{spl}[\hs]$, namely the initial conditions $x(-\infty)=\dot x(-\infty)=0$, allows positive frequency modes at the final time only.

It is worthwhile noting the difference between the schemes, obtained for harmonic oscillators in cases of the two different signs in Eq. \eq{ctps}. The trajectories $x^+(t)$ and $x^-(t)$ are independent in the action \eq{cctpacte}, and they are coupled by the final condition \eq{ctpfc} only. In the limit $t_i\to-\infty$, $t_f\to\infty$ the influence of the final condition on the trajectories becomes weak at finite $t$, and it may have a finite impact only with the choice of the sign $-$ in Eq. \eq{ctps} because the uncorrelated action \eq{cctpacte} is vanishing as $x^+(t)-x^-(t)\to0$. In more precise terms, the denominator on the right hand side of Eq. (A7) in Ref. \cite{effthpch} is $\ord{\dt}$. If the sign $+$ is chosen in Eq. \eq{ctps} then the action is finite in the limit $x^+(t)-x^-(t)\to0$, the denominator is $\ord{\dt^0}$ in Eq. (A7) and the trajectories $x^+(t)$ and $x^+(t)$ decouple. Therefore the sign $-$ should be used  in Eq. \eq{ctps} to keep the CTP copies coupled by the final condition \eq{ctpfc} in the limit $t_i\to-\infty$, $t_f\to\infty$.

The generator functional, $W[\hj]$, of an interacting system can be found by solving the equation of motion by iteration, which gives a formal functional power series in $\hj$, and each contribution can conveniently be represented by tree-graphs. Nontrivial initial conditions, $x(t_i)=x_i$, $\dot x(t_i)=v_i$ can be taken into account by starting the Green functions, corresponding to $x_i=v_i=0$, carrying out the shift $x\to x+x_i$, and using  the source $j(t)\to j(t)-mv_i\delta(t-t_i)$. If the initial velocity follows a probability distribution then the Green functions are given by the generator functional
\bea\label{clthev}
\overline{W}[\hj]&=&\sum_{n=0}^\infty\frac1{n!}\sum_{\sigma_1,\ldots,\sigma_n}\sigma_1\cdots\sigma_n\int dt_1\cdots dt_nD^{\sigma_1,\ldots,\sigma_n}(t_1,\ldots,t_n)\nn
&&\times\overline{[j^{\sigma_1}(t_1)-mv_i\delta(t_1-t_i)]\cdots[j^{\sigma_n}(t_n)-mv_i\delta(t_n-t_i)]},
\eea
where the bar stands for the expectation value.

\subsection{Effective action}\label{ceffths}
The traditional action principle, based an holonomic forces and boundary 
conditions in time, has been extended for semiholonomic forces and boundary conditions in Sec. \ref{semihs}, and for holonomic forces and initial conditions in Sec. \ref{initcs}. We now complete the construction of the action principle for effective theories by (i) showing that the effective forces within a system, involving holonomic forces are semiholonomic and (ii) generalizing the scheme of Sec. \ref{initcs} for semiholonomic forces.

The argument is followed within our simple model of Sec. \ref{intics}, whose action is written in the form $S[x,y]=S_s[x]+S_e[x,y]$. The effective action is easy to find by applying the method of Sec. \ref{intics} within the formalism of Sec. \ref{initcs} and the resulting effective action is $S_{eff}[\hx]=S[\hx,\hat y[\hx]]$, where the environment trajectory, $\hy=\hy[\hx]$, is obtained by solving the environment equations of motion for a general system trajectory, $\hx$. The environment initial conditions are by now built into the effective action, and point (ii) is completed. The argument of point (i) consists of simply noting that the CTP symmetry of the action $S[\hx,\hat y]$ assures us that $S_{eff}[\hx]$ displays CTP symmetry, as well, thereby describing semiholonomic forces. In other words, the form of the CTP action with semiholonomic forces is stable against the elimination of degrees of freedom. 

Let us have a closer look on the way semiholonomic forces appear in the effective theory. By writing the CTP action as $S[\hx,\hy]=S[x^+,y^+]-S[x^-,y^-]+S_{BC}[\hx,\hy]$, where the last term is infinitesimal and complex the form
\be\label{ceffe}
S_{eff}[\hx]=S[x^+,y^+[\hx]]-S[x^-,y^-[\hx]]+S_{BC}[\hx,\hy[\hx]]
\ee
is found where $\hy=\hy[\hx]$ is the solution of the equation of motion,
\be\label{envelim}
\fd{S[\hx,\hy]}{\hy}=0,
\ee
and the environment initial conditions. The CTP symmetry assures us that trajectory $\hy[\hx]$ is given in terms of a single functional, $y_2[x^+,x^-]$, as $y^\pm=y_2[x^\pm,x^\mp]$. We write Eq. \eq{ceffe} as
\be
S_{eff}[\hx]=S_s[x^+]-S_s[x^-]+S_{infl}[\hx]+S_{BC}[\hx],
\ee
where the impact of the environment on the system is summarized by the classical analogue of the influence functional \cite{feynman},
\be\label{cinflf}
S_{infl}[\hx]=S_e[x^+,y_2[\hx]]-S_e[x^-,y_2[\tau\hx]],
\ee
which transforms according to Eq. \eq{ctpsym} under CTP conjugation. Note that the imaginary part of the classical CTP action always remains infinitesimal, and it only encodes the initial conditions for the Green functions.

The finite, real part of the equation of motion,
\be\label{rceeom}
\fd{S_s[x^\pm]}{x^\pm}\pm\fd{S_{infl}[\hx,\hy[\hx]]}{x^\pm}=0,
\ee
shows that the nonconservative, open features of the effective dynamics come from the influence functional. It is advantageous to introduce the Keldysh parametrization \cite{keldysh}, $x^\pm=x\pm x^d/2$, at this point. The on-shell condition, $x^d=0$, is automatically satisfied by the solution and makes it sufficient to keep the $\ord{x^d}$ part of the real part of effective action, 
\be
\Re S_{eff}[x,x^d]=x^d\left(\fd{S_s[x]}{x}+\fd{S_{infl}[\hx]}{x^+}_{|x^+=x^-=x}\right)+\ord{x^{d2}}.
\ee
This form of the action indicates that the equation of motion, arising from the variation of $x$ and $x^d$, yields a trivial equation, $\ord{\epsilon}=0$, and 
\be
\fd{S_s[x]}{x}+\fd{S_{infl}[\hx]}{x^+}_{|x^+=x^-=x}=0,
\ee
respectively. The influence functional indeed describes semiholonomic forces, as mentioned in point (i) above.

Consider the effective Lagrangian 
\be\label{newtonfr}
L=\frac{m}2(\dot x^+)^2-U(x^+)-\frac{m}2(\dot x^-)^2+U(x^-)+\frac{k}2(x^-\dot x^+-x^+\dot x^-),
\ee
as a simple example. The equation of motion \cite{bateman}, 
\be
m\ddot x^\pm=-k\dot x^\mp-U'(x^\pm),
\ee
shows clearly that the semiholonomic force is represented by the passive CTP copy.

\subsection{Coupling of CTP copies}\label{ccctpcs}
The arbitrary system trajectory is not necessarily on-shell and the trajectory $y_2[\hx]$, obtained by solving Eq. \eq{envelim}, contains more information than the true ``conditional'' environment  trajectory, $y_1[x]=y_2[x,x]$, realized by a physical, on-shell system trajectory. To understand the role of this extra information that seems difficult to fit into Newton's mechanics we separate the single and double CTP copy contributions by writing $S_{infl}[\hx]=S_{1i}[x^+]-S_{1i}[x^-]+S_2[\hx]$. The comparison with Eq. \eq{cinflf} suggests the definitions $S_{1i}[x]=S_e[x,y_1[x]]$ and $S_2[\hx]=\Delta S_e[\hx]-\Delta S_e[\tau\hx]$, with
\be\label{deltask}
\Delta S_e[\hx]=S_e[x^+,y_2[\hx]]-S_e[x^+,y_1[x^+]].
\ee

The building up of the effective interactions can be understood as a two step process: First the system acts on its environment, and after that the modified environment acts back on the system. The first step, the modification of the environment dynamics by a fixed system trajectory, can generate holonomic forces only. The second step produces both holonomic and non-holonomic forces. The action $S_{1i}[x]$ contains the holonomic contributions of an environment, whose dynamics is based on initial conditions and the effective action of \eq{neffa}, $S_{eff}[x]=S_s[x]+S_{1i}[x]$. The problem with this effective action is that it cannot give account of nonholonomic forces. They are left to be represented by the coupling of the two CTP copies, collected in $S_2[\hx]$. Their impact on the conservation laws is discussed briefly in Appendix \ref{nothera}. 

It is interesting to check the separation of the effective interactions, mentioned above in a simple soluble harmonic system, defined by the action \cite{rubin,senitzky,ford,ullersma,caldeira}
\be\label{toyactc}
S[\hx,\hy;\hj]=\hf\hx\hD_s^{-1}\hx+\hf\hy\hD_e^{-1}\hy-\hx\hs(g\hy+\hj),
\ee
$\hj=(j,j)$ being an external, physical source. The environment trajectory,
\be
\hy(t)=\int dt'\hD_e(t,t')\hs\hx(t'),
\ee
is found by solving the environment equation of motion and the effective action turns out to be
\be
S_{eff}[\hx]=\hf\hx(\hD_s^{-1}-\hs\hD_e\hs)\hx.
\ee
The semiholonomic forces acting on the environment are described by
\be
y_2(t)=g\int dt'[D^n_e(t,t')x^+(t')+D^f_e(t,t')x^-(t')],
\ee
and the holonomic forces give
\be
y_1(t)=g\int dt'D^r_e(t,t')x(t').
\ee
Finally, one finds $S_{1i}[x]=-g^2xD_e^nx/2$ and $S_2[\hx]=-g^2x^+D^fx^-$.

An important realization of this model is where $x$ describes a point charge and its environment, $y$, is the electromagnetic field \cite{effthpch}. The effective theory is an action-at-a-distance model where the holonomic forces are mediated by the near field and are taken into account by $S_{1i}[x]$ \cite{schwa,tetr,fokk,wheel}. The novelty of the CTP scheme is the incorporation of the semiholonomic far field interactions in the action principle.

\section{Quantum Dynamics}\label{qctpds}
There is an obvious difference between classical and quantum mechanics from the point of view of the auxiliary conditions: We need two data for each degree of freedom in Newtonian mechanics but it is enough to provide the initial wave function in quantum mechanics and the initial conditions of classical mechanics are recovered in quantum mechanics at the level of averages only. 

We note in passing that the other kind of classical auxiliary condition, the specification of the initial and final coordinates, can be made in an exact manner in quantum mechanics: this is what happens when transition amplitudes are used in quantum mechanics. The projection of the current state of the system on a prescribed vector is made possible by the linear superposition principle. This is what happens when experimentalists choose certain kinematical cuts to detect particles with restricted energy momentum. In the usual presentation of scattering processes in classical physics one assumes an incoming, homogeneous particle flux, a classical realization of the linear superposition. 

The first order nature of the Schr\"odinger equation renders the initial condition problem trivial in quantum mechanics, and one simply considers the expectation value $\la A(t)\ra=\la\Psi(t)|A|\Psi(t)\ra$. Do we really need the extension of the well known formalism of quantum mechanics to semiholonomic forces if there is no difficulty in letting the environment follow an unconstrained time evolution? It is remarkable that though we do not need the reduplication of the degrees of freedom to calculate expectation values, nevertheless we have already got it, in the form of using bra and ket, both representing the same state.

\subsection{Semiholonomic forces, the density matrix, and the CTP formalism}\label{semiholq}
The naive quantization of the system with Hamiltonian \eq{ctpham}, with $H(x,p)=p^2/2m+U(x)$, leads to the extended Schr\"odinger equation,
\be
i\hbar\partial_t\psi(x^+,x^-,t)=\left[-\frac{\hbar^2}{2m}\partial^2_{x^+}+U(x^+)+\frac{\hbar^2}{2m}\partial^2_{x^-}-U(x^-)\right]\psi(x^+,x^-,t),
\ee
which looks like the equation of motion for the density matrix, $\psi(x,x',t)\to\rho(x,x',t)$. The reduplication of the degrees of freedom to accommodate initial condition problems and semiholonomic forces in classical mechanics leads naturally to mixed states and their representation by the density matrix, shedding some new light on the Gleason theorem \cite{gleason}. 

The transformation under local $U(1)$ gauge transformation, 
\be
\psi(x^+,x^-,t)\to e^{i[\alpha(x^+)-\alpha(x^-)]}\psi(x^+,x^-,t),
\ee
which is compatible with a possible external electromagnetic field, shows that $x^+$ and $x^-$ correspond to a bra and a ket, $\rho(x^+,x^-,t)=\la x^+|\rho(t)|x^-\ra$. One may introduce at this point two Hilbert spaces, ${\cal H}^\pm$, where the canonical operators, $x^\pm$ and $p^\pm$, act \cite{umezawa}, and the ``wave function'' alias density matrix, becomes an element of the Louville space \cite{shmutz}, ${\cal H}^+\otimes{\cal H}^-$, and the expectation values become linear in the CTP ``wave function'', $\rho(x^+,x^-,t)$, underlying the difference of the scalar product in the Hilbert space ${\cal H}^\pm$ of pure states and in the Liouville space of operators.

Rather than following this way of thinking we return to the equivalent, standard scheme of quantum mechanics and define the generator functional of connected Green functions of the coordinate \cite{schw},
\be\label{ctpgfunc}
e^{\ih W[\hj]}=\Tr[U(t_f,t_i;j^+)\rho(t_i)U^\dagger(t_f,t_i;j^-)],
\ee
to facilitate the perturbation expansion of $x$-dependent observables, where the reduplication of the degrees of freedom comes from the double appearance of the time evolution operator. This equation is written in the Heisenberg representation, where 
\be\label{ctpgfncto}
U(t_f,t_i;j)=T[e^{-\ih\int_{t_i}^{t_f}dt'[H(t)-x(t)j(t)]}]
\ee
and $T$ denotes the time ordering. It is advantageous to introduce an extended time ordering, $\bar T$, which is the usual or the antitime ordering for factors acting on ${\cal H}^+$ or ${\cal H}^-$, respectively, and places the operators of ${\cal H}^-$ left of those of ${\cal H}^+$. This allows us to write the generator functional \eq{ctpgfunc} as
\be
e^{\ih W[\hj]}=\Tr\left(\bar T\left[e^{\ih\int_{t_i}^{t_f}dt[H^-(t)-x^-(t)j^-(t)]}e^{-\ih\int_{t_i}^{t_f}dt[H^+(t)-x^+(t)j^+(t)]}\right]\rho(t_i)\right),
\ee
where the Hamiltonian, $H^\pm$, is constructed of the operators $x^\pm$ and $p^\pm$. The achievement of this formalism is that the Wick theorem applies for this extended time ordered product and Feynman rules can be derived.

Note that the unitarity of $U(t_f,t_i;j)$ is relevant within the subspace that is spanned by states, visited by the system in the presence of the source $j(t)$. It is expressed by the conservation of the norm, $W[j,j]=0$, and the $t_f$ independence of expectation values, calculated at $t<t_f$: cf. the remark made after Eq. \eq{ctpsym}.

The path integral representation of the generator functional, \eq{ctpgfunc}, is
\be\label{ctpqmgf}
e^{\ih W[\hj]}=\int_{x^+(t_f)=x^-(t_f)}D[\hx]e^{\ih S[\hx]+\ih\hj\hs\hx}\rho(x^+(t_i),x^-(t_i),t_i),
\ee
where the final condition \eq{ctpfc} is imposed by the trace in Eq. \eq{ctpgfunc}, evaluated in coordinate representation. The connected Green functions are defined by Eq. \eq{ctpgfnctd}: in particular the harmonic oscillator of the Lagrangian \eq{holagr} yields $W_0[\hj]=\hj\hD_0\hj/2$, where the propagator, 
\be
i\hbar\hD(t,t')=\begin{pmatrix}\la0|T[x(t)x(t')]|0\ra&\la0|x(t')x(t)|0\ra\cr\la0|x(t)x(t')|0\ra&\la0|T[x(t')x(t)]|0\ra^*\end{pmatrix},
\ee
is given by Eq. \eq{hogfnct}. We note finally that the contact of our harmonic oscillator with a heat bath modifies the free propagator in an additive manner \cite{ed},
\be
\hD_{th}(\omega)=\hD(\omega)-\frac{i}m2\pi\delta(\omega^2-\Omega^2)n(\Omega)\begin{pmatrix}1&1\cr1&1\end{pmatrix},
\ee
where $n(\omega)=1/(e^{\beta\hbar\omega}-1)$.

Large systems are handled within the framework of quantum field theory. The generator functional for the connected system Green function can be found by extending the expression \eq{ctpqmgf} to quantum fields with the perturbative vacuum as the initial condition, $\rho(t_i)=|0_p\ra\la0_p|$,
\be\label{ctpcggf}
e^{\ih W[\hj]}=\la0_p|\bar T\left[e^{\ih\int dtH^-(t)-\ih\phi^-j^-]}e^{-\ih\int dxH^+(t)+\ih\phi^+j^+}\right]|0_p\ra,
\ee
where $H^\pm(t)$ is the energy, constructed by the fields $\pm$, and the scalar product of functions over the space-time stands for $fg=\int dxf(x)g(x)$. The usefulness of this functional hinges on the assumption that the true vacuum develops from the perturbative one during the time evolution and all physically relevant initial states can be reached at $t=0$ by guiding the system adiabatically with a physical external source, $\hj=(j_{ph},j_{ph})$, by starting at $t_i=-\infty$. The path integral expression of this functional is
\be
e^{\ih W[\hj]}=\int D[\hphi]e^{\ih S[\phi^+]-\ih S[\phi^-]+\ih S_{BC}[\hphi]+\ih\hphi\hs\hj},
\ee
where the integration is over the CTP copies, $\hphi=(\phi^+,\phi^-)$, satisfying the final condition \eq{ctpfc}, $\phi^+(t_f,\v{x})=\phi^-(t_f,\v{x})$, and the initial density matrix is suppressed in the condensed notation. We take the limit $t_i\to-\infty$, $t_f\to\infty$, and use
\be\label{bcact}
S_{BC}[\hphi]=\frac{i\epsilon}2\int dx[(\phi^+(x))^2+(\phi^-(x))^2]+\frac\epsilon\pi\int dxdx'\frac{\phi^+(x)\phi^-(x')}{x^0-x'^0+i\epsilon},
\ee
cf. \eq{acctpact}.

For a free field, $S_0[\phi]=\phi K_0\phi/2$, one finds the CTP action
\bea\label{freectpact}
S_0[\hphi]&=&\hf\phi^+K_0\phi^+-\hf\phi^-K_0\phi^-+S_{BC}[\hphi]\nn
&=&\hf\hphi\hat K_0\hphi,
\eea
which leads to the free generator functional, 
\be\label{freectgf}
e^{\ih W_0[\hat k]}=\int D[\hphi]e^{\ih S_0[\hphi]+\ih\hat k\hs\hphi}=e^{-\frac{i}{2\hbar}\hat k\hD_0\hat k}.
\ee
The free CTP propagator,
\be
i\hbar\hD_0(x,y)=\begin{pmatrix}\la0|T[\phi(x)\phi(y)]|0\ra&\la0|\phi(y)\phi(x)|0\ra\cr\la0|\phi(x)\phi(y)|0\ra&\la0|T[\phi(y)\phi(x)]|0\ra^*\end{pmatrix},
\ee
is easiest to calculate in the operator formalism. In the case of a free field with mass $m$ we have $K_0=-\Box-m^2$, and the standard steps lead to the Fourier transform
\bea
\hD_0(k)&=&\int dxe^{ikx}\hD_0(x),\nn
&=&{\cal C}_3[D^n_0(k),D^f_0(k),D^i_0(k)],
\eea
where 
\bea\label{ctpscpr}
D_0^n(k)&=&P\frac1{k^2-m^2},\nn
D_0^f(k)&=&-i\pi\sign(k^0)\delta(k^2-m^2),\nn
D_0^i(k)&=&-\pi\delta(k^2-m^2),
\eea
in the limit $t_i\to-\infty$, $t_f\to\infty$ cf. Eq. \eq{hogfnct}, after having set $\hbar=c=1$. It is easy to verify by inversion that $\hat K_0=\hs{\cal C}_3[K^n_0,K^f_0,K^i_0]\hs$ with 
\bea
K_0^n(k)&=&k^2-m^2,\nn
K_0^f(k)&=&i\sign(k^0)\epsilon,\nn
K_0^i(k)&=&\epsilon,
\eea
cf. Eq. \eq{hoigfnct}.

There might be a heat or particle reservoir that can be taken into account perturbatively by the modification
\be\label{ctpscprn}
\hD_{res}(k)=\hD_0(k)-i2\pi\delta(k^2-m^2)n(k)\begin{pmatrix}1&1\cr1&1\end{pmatrix}
\ee
of the propagator where the occupation number,
\be
n(k)=\frac{\Theta(-k^0)}{e^{\beta(\epsilon(\v{k})+\mu)}-1}+\frac{\Theta(k^0)}{e^{\beta(\epsilon(\v{k})-\mu)}-1},
\ee
may have $\mu\ne0$ for the conserved particle number. The inverse can easily be calculated, and one finds
\be\label{envfefa}
\hat K_{res}(k)=\hat K_0(k)+2in\epsilon\begin{pmatrix}1&-1\cr-1&1\end{pmatrix},
\ee
which can be written as $\hat K_{res}=\hat K_0+{\cal C}_3[0,0,2n\epsilon]$, the $\ord{\epsilon}$ part modifies the occupation number in the initial state.

Finally we mention a few properties of the two-point function that remain valid for any bosonic local operator. (i) The identity
\be\label{ctpid}
T[\phi(x)\phi(y)]+T^*[\phi(x)\phi(y)]=\phi(x)\phi(y)+\phi(y)\phi(x)
\ee
cf. Eq. \eq{cctpid}, where $T^*$ denotes the anti timeordering, implies the block structure $\hD={\cal C}_3[D^n,D^f,D^i]$. (ii) The states, contributing to the spectral function,
\be\label{spfnct}
iD^{-+}(k)=\sum_n\la0|\phi(-k)|n\ra\la n|\phi(k)|0\ra,
\ee
have positive energy therefore $D^{-+}(k)=0$ for $k^0\le0$, cf. the remark made after Eq. \eq{ctpfcf}. The resulting relation, 
\be\label{posencond}
iD^i(k)=\mr{sign}(k^0)D^f(k),
\ee
reduces the number of independent functions of the propagator to two, $\hD(k)={\cal C}_2[D^n(k),D^f(k)]$, where
\be
{\cal C}_2[D^n(k),D^f(k)]
=\begin{pmatrix}D^n(k)+\mr{sign}(k^0)D^f(k)&-2\Theta(-k^0)D^f(k)\cr2\Theta(k^0)D^f(k)&-D^n(k)+\mr{sign}(k^0)D^f(k)\end{pmatrix},
\ee
and $\hat K(k)=\hD^{-1}(k)=\hs{\cal C}_2[K^n(k),K^f(k)]\hs$. (iii) Assuming that the norm of the states that contribute to the spectral sum \eq{spfnct} is positive, we arrive at the bound 
\be\label{dibound}
D^i(k)\le0.
\ee
(iv) The diagonal blocks are given by the Feynman propagator, $\Im D^{++}(x,y)=D^i(x,y)$ is on shell but $\Re D^{++}(x,y)=D^n(x,y)$ is off shell due to the time ordering in the Feynman propagator. The off-diagonal blocks, $D^{\pm\mp}$, are given by the Wightmann function and are on shell.

\subsection{Effective theories}\label{qeffths}
We now split the full system into an observed system and its environment, described by the fields $\phi(x)$ and $\psi(x)$, respectively, which are supposed to obey the dynamics of the action $S[\phi,\psi]=S_s[\phi]+S_e[\phi,\psi]$. The generator functional for the connected system Green functions,
\be
e^{\ih W[\hj]}=\int D[\hphi]D[\hpsi]e^{\ih S_s[\phi^+]-\ih S_s[\phi^-]+\ih S_e[\phi^+,\psi^+]-\ih S_e[\phi^-,\psi^-]+\ih S_{BC}[\hphi]+\ih S_{BC}[\hpsi]+\ih\hphi\hs\hj},
\ee
can be simplified by integrating over the environment variables,
\be\label{ctpefgf}
e^{\ih W[\hj]}=\int D[\hphi]e^{\ih S_{eff}[\hphi]+\ih S_{BC}[\hphi]+\ih\hphi\hs\hj},
\ee
where the effective action,
\be\label{qinffl}
S_{eff}[\hphi]=S_s[\phi^+]-S_s[\phi^-]+S_{infl}[\hphi],
\ee
contains the influence functional \cite{feynman},
\be\label{inflqft}
e^{\ih S_{infl}[\hphi]}=\int D[\hpsi]e^{\ih S_e[\phi^+,\psi^+]-\ih S_e[\phi^+,\psi^-]+\ih S_{BC}[\hpsi]}.
\ee

The perturbation expansion provides a general scheme to calculate the effective action for weakly coupled theories where $S_e[\phi,\psi]=S_{e0}[\psi]+S_{ei}[\phi,\psi]$, $S_{e0}[\psi]$ being the free action. The perturbation series is defined by expanding the exponential function in the interactions in the equation
\be\label{pertseries}
e^{\ih S_{infl}[\hphi]}=e^{\ih S_{ei}[\phi^+,\frac\hbar{i}\fd{}{k^+}]-\ih S_{ei}[\phi^-,\frac\hbar{i}\fd{}{k^-}]}e^{\ih W_0[\hat k]},
\ee
where $W_0[\hat k]$ is the free environment generator functional \eq{freectgf} constructed with the help of the free action, $S_{e0}[\psi]$. The coefficients of the powers of $\hphi$ are given as the sum of environment Feynman graphs where the external legs are represented by the factors of $\hphi$, the $\psi$ dependence of the terms in $S_{ei}[\phi,\psi]$ define the vertices, and the lines stand for the environment propagator. 

\begin{figure}[ht]
\includegraphics[scale=0.2]{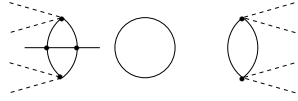}
\caption{An $\ord{g_1^2\lambda^4}$ graph contributing to $\la0|T[\phi(x)\phi(y)]|0\ra$. The circle represents the initial density matrix, $\rho(t_i)=|0_p\ra\la0_p|$, the continuous and the dashed  lines stand for the $\phi$ and the $\psi$ propagators, respectively. The part left (right) from the circle belong to $U$ ($U^\dagger$). One has $t=t_f$ at the left and right ends of the graph where the lines reaching this time from $U$ and $U^\dagger$ are joined.}\label{idgr}
\end{figure}

We continue the discussion with a generic model, defined by the Lagrangian
\be\label{sclagr}
L=\hf\partial_\mu\phi\partial^\mu\phi-\frac{m^2}2\phi^2-\frac{g_1}{4!}\phi^4+\hf\partial_\mu\psi\partial^\mu\psi-\frac{M^2}2\psi^2-\frac{g_2}{4!}\psi^4-\frac\lambda4\phi^2\psi^2.
\ee
A CTP Feynman graph, eg. the one shown in Fig. \ref{idgr}, has two parts that are separated by the circle, representing the initial density matrix. We place the two CTP copies on two different sides of the circles: Left and right are the lines and vertices that belong to $U$ and $U^\dagger$ in \eq{ctpgfunc}, respectively. The lines $D^{\pm\pm}$ represent the propagation of an excitation, controlled by $\psi^+$ or $\psi^-$, and are positioned at one side of the circle. The new feature of a CTP graph, the line $D^{\pm\mp}$, connects the two sides of the circle. To find its physical interpretation let us write the trace of the generator functional \eq{ctpgfunc} as a sum over a basis at the final time,
\be\label{resident}
e^{\ih W[\hj]}=\sum_n\la n|U(t_f,t_i;j^+)\rho(t_i)U^\dagger(t_f,t_i;j^-)|n\ra.
\ee
Such a representation of the trace has been used in the spectral function \eq{spfnct}, too, and it shows that the line $G^{\pm\mp}$ always represents a $\psi$ excitation in the final state. 

It is useful to separate the CTP graphs into three classes: A graph is called homogeneous if all external legs and vertices belong to the same CTP copy. The inhomogeneous graphs have all external legs in the same CTP copy but their vertices can be found in both copies. Finally, if both CTP copies can be found among the external legs then we talk a genuine CTP graphs. The characterization of the first and the third classes is easy: A homogeneous graph is obviously the same as in the traditional, non-CTP quantum field theory, based on transition amplitudes and a genuine CTP graph describes a process where the final state contains $\psi$ particles. 

The second class, the inhomogeneous graphs are vanishing if the initial state is the vacuum. In fact, consider those internal lines of a connected component of the graph that connect two different CTP copies, cf. Fig. \ref{idgr}. Let us count the frequency of these lines in the same direction, say $+\to-$, where we find a multiplicative factor $D^{-+}(p)$ for each line. These lines represent the positive energy final state, contributing to the trace in the generator functional \eq{ctpcggf}, e.g., four $\psi$ particles in the case of the graph of Fig. \ref{idgr}. They are on shell and the Heaviside function in the free propagator \eq{ctpscpr} assures their positive energy in the final states. Since all external legs belong to the same CTP copy, the sum of the frequencies of the internal lines in between the CTP copies is zero. Hence there is at least one negative frequency and the corresponding free propagator is vanishing. In other words, there are no on-shell excitations with vanishing energy. If there are states in the Fock space with lower energy than the initial state then the inhomogeneous graphs may be nonvanishing. This happens when the system is attached to a heat or particle reservoir and the second term in the right hand side of Eq. \eq{ctpscprn} contains no Heaviside function.

It is instructive to see how the exact interacting propagator is obtained by the Schwinger-Dyson resummation method when the system is attached to a reservoir. We write the inverse propagator as $\hat K=\hat K_0-\hs\hat\Pi\hs$, where both the free inverse propagator, $\hat K_0={\cal C}_2[K_0^n,K_0^f]+{\cal C}_3[0,0,\Delta K^i]$, and the self energy, $\hat\Pi={\cal C}_2[\Pi^n,\Pi^f]+{\cal C}_3[0,0,\Delta\Pi^i]$, can be written as the sum of the vacuum contribution, which satisfies the constraint \eq{posencond}, and the rest. The inversion is carried out by the help of eqs. \eq{ctppropinv}, and the exact Feynman propagator,
\bea\label{sdresprop}
D^{++}&=&D^a(K^n-iK^i)D^r\nn
&=&\frac1{K_0^n+iK_0^i-\Pi^n-i\Pi^i}+\frac{i(\Delta\Pi^i-\Delta K^i)}{(K_0^n+iK_0^i-\Pi^n-i\Pi^i)(K_0^n-iK_0^i-\Pi^n+i\Pi^i)},
\eea
is found where $K_0^i$ and $\Pi^i$ are given by Eq. \eq{posencond}. In the absence of the reservoir $\Delta K^i=\Delta\Pi^i=0$ and $\Pi^{++}$, being the sum of homogeneous graphs, agrees with the self-energy of the traditional quantum field theory formalism and $D^{++}$ is identical to the traditional Feynman propagator. In the presence of the reservoir the second line of \eq{sdresprop} gives the Feynman propagator in a form where the positive energy intermediate state contributions are collected in the first term. The second, imaginary term is well defined and finite in the kinematical regime $\Pi^i\ne0$ where there are on-shell, positive energy asymptotic collective excitations with the same quantum number as $\phi(x)$. 

The pinching singularities \cite{altherrs}, generated by poles with infinitesimal imaginary parts on both sides of the real frequency axis, make the second term ill defined for $\Pi^i=0$. If there are no on shell, negative energy states of the environment available that can contribute to the self-energy and both $\Pi^i$ and $\Delta\Pi^i$ are vanishing, then the second term is well defined and finite, $\epsilon/\epsilon$  \cite{altherr,baier,lebellac,niegawa,dadic}. The imaginary part of the propagator diverges for $\Pi^i=0$ and $\Delta\Pi^i\ne0$. But the perturbation expansion is not a reliable tool in this case. In fact, let us check the clusterization property of the propagator, 
\be\label{clust}
\la0|T[\phi(x)\phi(y)]|0\ra\to\la0|\phi(x)|0\ra\la0|\phi(y)|0\ra,
\ee
for $x^0\to y^0$ and $|\v{x}-\v{y}|\to\infty$ by carrying out a Fourier transformation of both sides in $x-y$. The right hand side, being given by the expectation value of Hermitian operators, is real. The left hand side  is symmetric with respect to the exchange $x\leftrightarrow y$ hence its real and imaginary parts in space-time belong to their real and imaginary parts in Fourier space. The limit \eq{clust} is violated because the left hand side has a diverging imaginary part.

We collect the contributions of all homogeneous and inhomogeneous graphs to the effective action into $S_h[\phi]$ and $S_{ih}[\phi]$, respectively. The single CTP copy contribution to the influence functional is therefore $S_{i1}[\phi]=S_h[\phi]+S_{ih}[\phi]$. We do not attach anymore a reservoir to the system: hence $S_{ih}[\phi]=0$ and $S_{i1}[\phi]$ agrees with the Wilsonian effective action of traditional quantum field theories. The full influence functional is of the form
\be
S_{infl}[\hphi]=S_{i1}[\phi^+]-S_{i1}^*[\phi^-]+S_2[\hphi],
\ee
where $S_2[\hphi]$ contains genuine CTP graphs only, representing the coupling between the CTP copies of the system and describing processes with $\psi$ particles in the final state. Therefore a part of the CTP effective action, $S_{i1}[\phi]$, includes the effective vertices of the traditional effective quantum field theories since their final state is the vacuum, but there are in addition effective interactions in $S_2[\hphi]$ that leave the environment in an excited state.

It is sometimes useful to express the effective action in the Keldysh parametrization, $\phi^\pm=\phi\pm\phi^d/2$,
\be\label{effackp}
S_{eff}[\phi,\phi^d]=S_1\left[\phi+\frac{\phi^d}2\right]-S^*_1\left[\phi-\frac{\phi^d}2\right]+S_2\left[\phi+\frac{\phi^d}2,\phi-\frac{\phi^d}2\right],
\ee
where $S_1[\phi]=S_s[\phi]+S_{i1}[\phi]$. The expansion is carried out around the expectation values, $\la0|\phi|0\ra=\bar\phi$ and $\la0|\phi^d|0\ra=0$, up to quadratic order,
\bea\label{eakp}
S_{eff}[\phi,\phi^d]&=&i\Im(2S_1+S_2)+2i\delta\phi\left(\fd{\Im S_1}{\phi}+\fd{\Im S_2}{\phi^+}\right)+\delta\phi^d\left(\fd{\Re S_1}{\phi}+\fd{\Re S_2}{\phi^+}\right)\nn
&&+\delta\phi^d\fdd{\Re S_2}{\phi^+}{\phi^-}\delta\phi+i\delta\phi\left(\fdd{\Im S_1}{\phi}{\phi}+\fdd{\Im S_2}{\phi^+}{\phi^+}+\fdd{\Im S_2}{\phi^+}{\phi^-}\right)\delta\phi\nn
&&+\frac{i}4\delta\phi^d\left(\fdd{\Im S_1}{\phi}{\phi}+\fdd{\Im S_2}{\phi^+}{\phi^+}-\fdd{\Im S_2}{\phi^+}{\phi^-}\right)\delta\phi^d,
\eea
where $S_1$ and $S_2$ are evaluated at $\phi^\pm=\bar\phi$. The invariance of the path integral \eq{ctpefgf} under the shift of the integral variable, $\hphi\to\hphi+\delta\hphi$, yields the equation of motion,
\be
\fd{S_1}{\phi}+\fd{S_2}{\phi^+}=0
\ee
satisfied on the level of matrix expectation value. 

A harmonic toy model is defined by the action
\be\label{ftoyact}
S[\hphi,\hpsi]=\hf\hphi\hat K_s\hphi+\hf\hpsi\hat K_e\hpsi-\hphi\hs(g\hpsi+\hj)
\ee
cf. Eq. \eq{toyactc}. The elimination of the environment gives
\bea\label{ftoyacts}
S_1[\phi]&=&\hf\phi(K^n_{eff}+iK^i_{eff})\phi,\nn
S_2[\hphi]&=&\phi^+(K^f_{eff}-iK^i_{eff})\phi^-,
\eea
with $\hat K_{eff}=\hat K_s-g^2\hs\hD_e\hs$. The effective action assumes the form
\be\label{effactoyf}
S_{eff}[\hphi]=\hf\left[\phi K^a_{eff}\phi^d-\phi^dK^r_{eff}\phi+\phi^diK^i_{eff}\phi^d\right]
\ee
in the Keldysh parametrization.

\subsection{Scattering}\label{scatts}
After outlining the general structure of effective CTP theories, we mention two applications briefly, the first being inclusive scattering processes. Consider, for instance, a two $\phi$-particle scattering, $\v{p}_{i1}+\v{p}_{i2}\to\v{p}_{f1}+\v{p}_{f2}$, in the model \eq{sclagr} and write the transition probability as the expectation value
\be\label{trprobsta}
P=\Tr[a(t_i,\v{p}_{i2})a(t_i,\v{p}_{i1})a^\dagger(t_f,\v{p}_{f2})a^\dagger(t_f,\v{p}_{i1})a(t_f,\v{p}_{f1})a(t_f,\v{p}_{f2})a^\dagger(t_i,\v{p}_{i1})a^\dagger(t_i,\v{p}_{i2})|0\ra\la0|].
\ee
The scattering process in an unobserved environment is always inclusive, and it leaves the environment in an unknown state. The calculation of \eq{trprobsta} starts with the application of the reduction formulas, applied to the generator functional \eq{ctpcggf} in the limit $t_f\to-\infty$, $t_f\to\infty$. If all asymptotic particles are extracted from the same time axis, then the stability of the vacuum, $|0\ra=U(t_f,t_i,0)|0\ra$, makes the resulting probability exclusive and equivalent to the result, found in traditional quantum field theory. To get the probability of the inclusive scattering process we have to extract the first and the second chains of four operators in Eq. \eq{trprobsta} from $U^\dagger$ and $U$, respectively \cite{scatt}.  In fact, if the trace in the generator functional \eq{ctpcggf} is calculated by summing over a basis as in Eq. \eq{resident}, then we obtain an inclusive scattering probability by summing over exclusive scattering processes with different $\psi$-particle content in the final state. 

\begin{figure}[ht]
\includegraphics[scale=0.2]{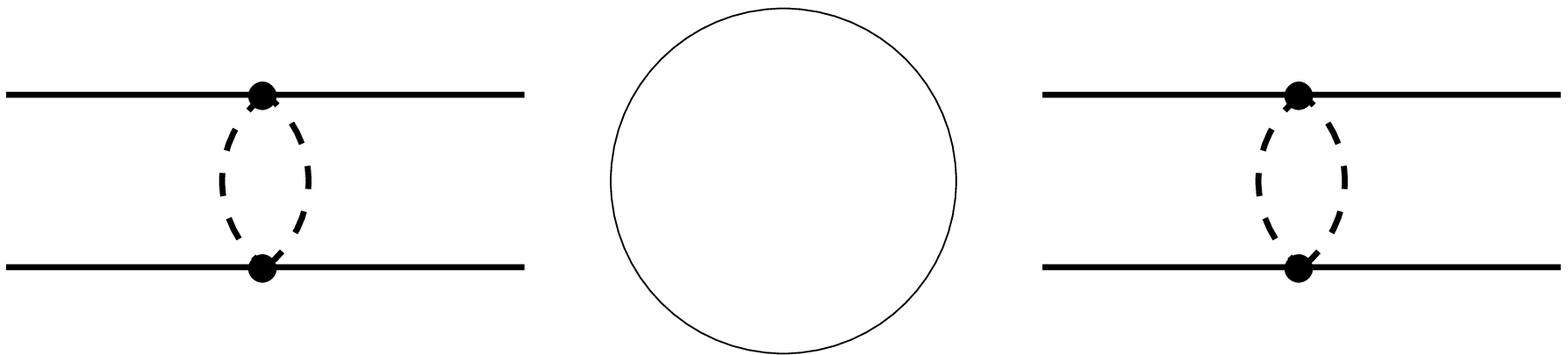}\\(a)\\
\includegraphics[scale=0.2]{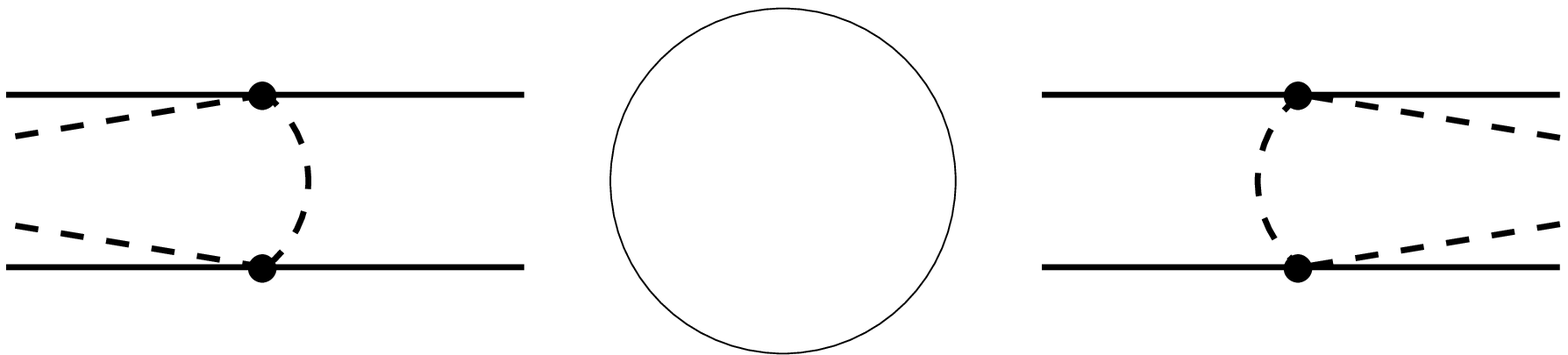}\\(b)\\
\includegraphics[scale=0.2]{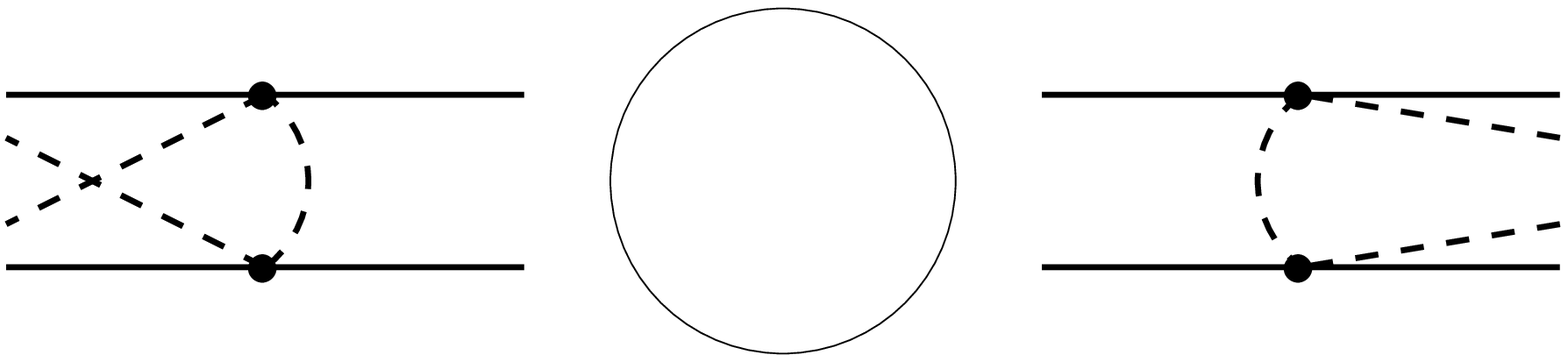}\\(c)\\
\includegraphics[scale=0.2]{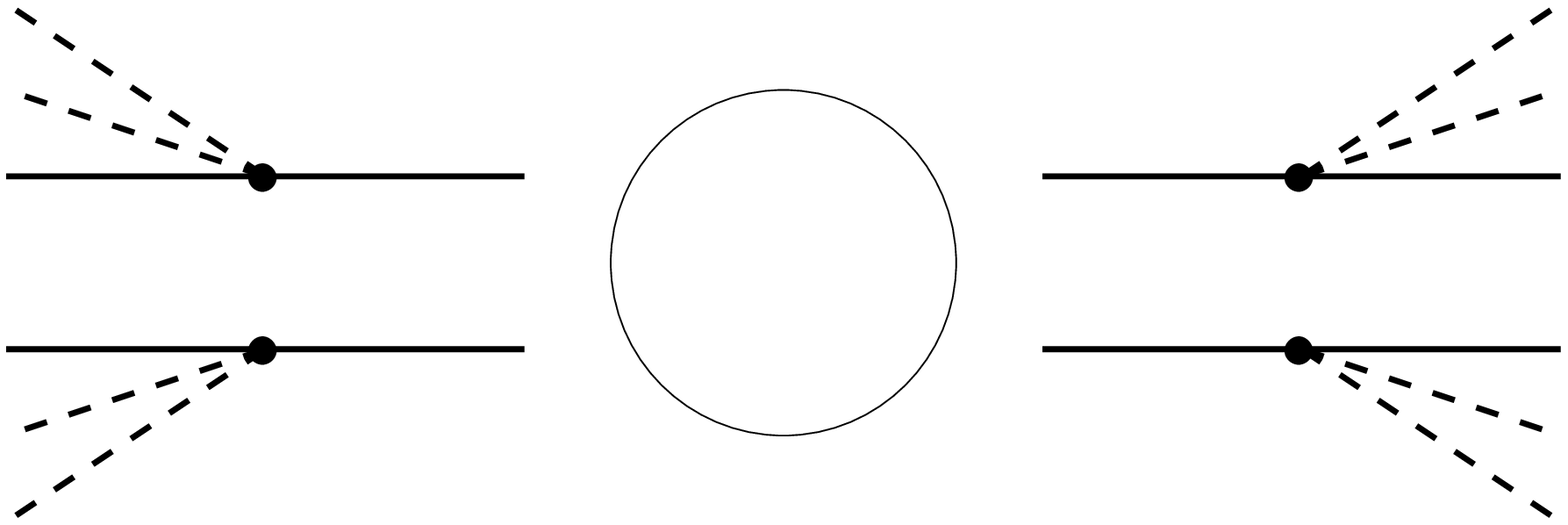}\\(d)\\
\includegraphics[scale=0.2]{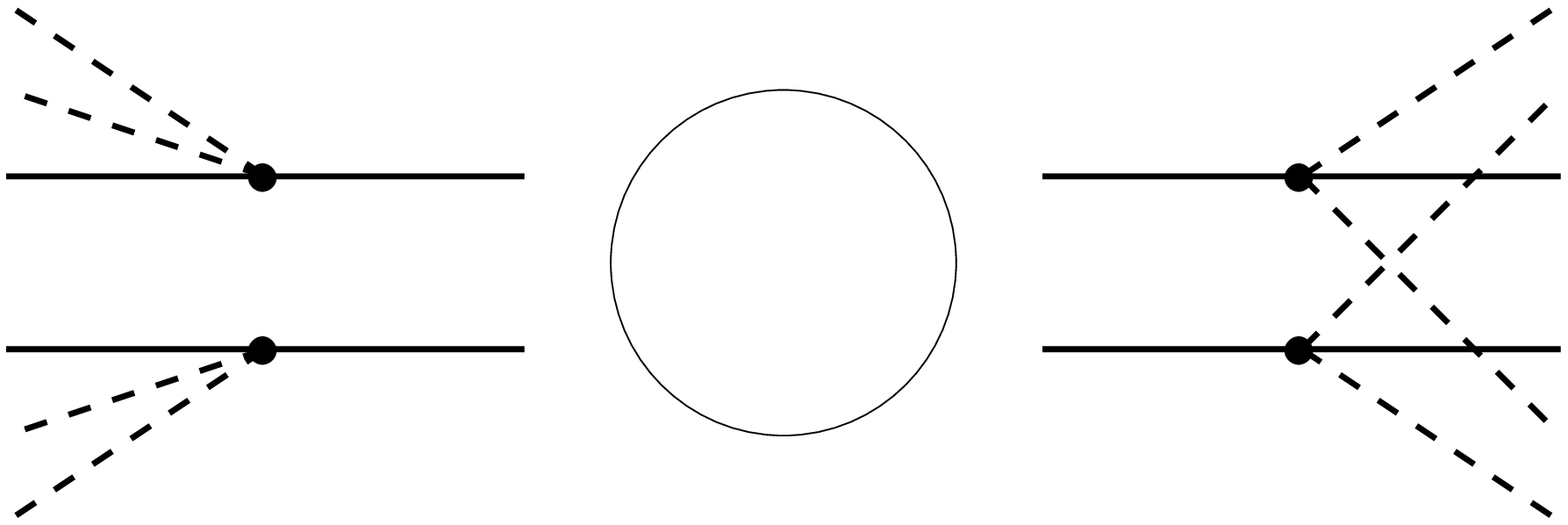}\\(e)\\
\includegraphics[scale=0.2]{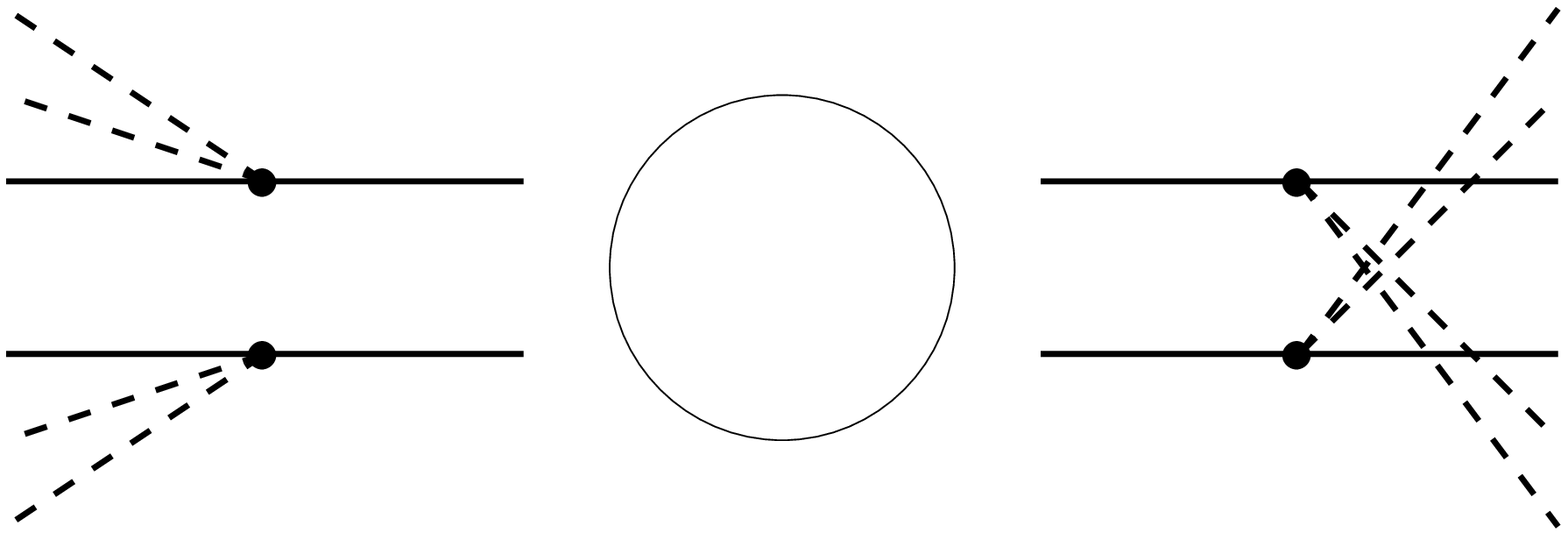}\\(f) 
\caption{$\ord{\lambda^4}$ graphs that contribute to the two $\phi$ particle nonforward scattering probability with $n_r$ real and $n_v$ virtual $\psi$ particles with $(n_r,n_v)$ given by (a) (0,2); (b)-(c) (2,1) (d)-(e)-(f) (4,0). (a) Exclusive scattering; (b)-(c) inclusive scattering; and (d)-(e)-(f) represent the contributions of relative states \eq{relst} with $n=2$ and $n=4$, respectively.}\label{scatt}
\end{figure}

The Fock-space vector, representing the scattered state, is of the form
\be
|\Psi\ra=\sum_{n=0}^\infty\int\frac{d^3p_1d^3p_2d^3q_1\cdots d^3q_n}{(2\pi)^{3(2+n)}}\Psi_n(\v{p}_1,\v{p}_2,\v{q}_1,\ldots,\v{q}_n)a^\dagger(\v{p}_1)a^\dagger(\v{p}_2)b^\dagger(\v{q}_1)\cdots b^\dagger(\v{q}_n)|0\ra,
\ee
where $a^\dagger(\v{q})$ and $b^\dagger(\v{q})$ stand for the creation operator of a $\phi$ and a $\psi$ particle, respectively. The system-environment entanglement will be followed by means of the relative state \cite{everett}, defined in Appendix \ref{relsta}. The relative state represents the conditional system state, assuming that the state of the environment particles is known,
\be\label{relst}
|R(\v{q}_1,\ldots,\v{q}_n)\ra=N(\v{q}_1,\ldots,\v{q}_n)\int\frac{d^3p_1d^3p_2}{(2\pi)^6}\Psi_n(\v{p}_1,\v{p}_2,\v{q}_1,\ldots,\v{q}_n)a^\dagger(\v{p}_1)a^\dagger(\v{p}_2)|0\ra,
\ee
where the normalization factor is defined by the equation
\be
\frac1{N^2(\v{q}_1,\ldots,\v{q}_n)}=\int\frac{d^3p_1d^3p_2}{(2\pi)^6}|\Psi_n(\v{p}_1,\v{p}_2,\v{q}_1,\ldots,\v{q}_n)|^2=P(\v{q}_1,\ldots,\v{q}_n).
\ee
The corresponding reduced density matrix of the scattered system is
\be\label{scdensm}
\rho=\sum_{n=0}^\infty\int\frac{d^3q_1\cdots d^3q_n}{(2\pi)^{3n}}|R(\v{q}_1,\ldots,\v{q}_n)\ra P(\v{q}_1,\ldots,\v{q}_n)\la R(\v{q}_1,\ldots,\v{q}_n)|.
\ee
Figure \ref{scatt} shows a few graphs that contribute to the transition probability. The inclusive scattering processes, depicted in Figs. \ref{scatt} (b)-\ref{scatt} (f) correspond to nontrivial relative states. The summation over the momenta of $\psi$ particles in the final states, possible environment excitations, in Eq. \eq{scdensm} gives account of the system-environment entanglement. 

It is an advantageous feature of the CTP scheme when applied to the calculation of scattering probability that it produces observable transition probabilities rather than amplitudes. In fact, the probability, being bounded by unitarity, includes automatically all contributions needed for the cancellation of collinear divergences between the virtual and real particles \cite{yennie}.

\subsection{Causality}\label{causals}
Another phenomenon where the CTP formalism offers a new insight is causality that we  considered here within the model \eq{sclagr} by inspecting the expectation value of the composite operator $\rho(x)=\phi^2(x)$ in the presence of an external classical potential, $u(x)$, coupled linearly to $\rho(x)$. 

In the traditional formalism one uses the generator functional
\bea
e^{\ih W_\rho[u]}&=&\la0|U(t_f,t_i;u)|0\ra\nn
&=&\int D[\phi]D[\psi]e^{\ih S[\phi,\psi]+\ih u\rho},
\eea
to calculate
\be
\la0|T[U(t_f,t_i;u)\rho(x)]|0\ra=\fd{W_\rho[u]}{u(x)}.
\ee
Note that this is an expectation value only if the vacuum is stable, $\la0|U(t_f,x^0;u)=\la0|$, which is usually satisfied if $u(y)=0$ for $y^0>x^0$. The leading order expression in $u$,
\be
\la0|T[U(t_f,t_i;u)\rho(x)]|0\ra=-\int dy D_\rho(x,y)u(y),
\ee
involves the composite operator Feynman propagator,
\bea
D_\rho(x,y)&=&\fdd{W_\rho[u]}{u(x)}{u(y)}_{|u=0}\nn
&=&\la0|T[\rho(x)\rho(y)]|0\ra.
\eea
Being symmetric under the exchange $x\leftrightarrow y$ the external potential acts forward and backward in time. This is a general feature of the Feynman propagator, the vacuum expectation value of a time ordered product: The relativistic quantum fields contain positive and negative frequency components that act forward (creation) or backward (annihilation) in time. The forward action in time results from the fixed initial condition, $|0\ra$, and the backward action owes its existence to the final condition, $\la0|$. 

If the final state is unknown, e.g., $u(y)\ne0$ for $y^0>x^0$ in the example above, then we need the CTP formalism to find expectation values. The generator functional
\bea
e^{\ih W_\rho[\hat u]}&=&\Tr[U(t_f,t_i;u^+)|0\ra\la0|U^\dagger(t_f,t_i;-u^-)]\nn
&=&\int D[\hphi]D[\hpsi]e^{\ih S[\phi^+,\psi^+]-\ih S^*[\phi^-,\psi^-]+\ih\hat u\hs\hat\rho}
\eea
gives two equivalent expressions for the expectation value,
\be\label{eqexpv}
\la0|\rho(x)|0\ra=\fd{W_\rho[\hat u]}{u^+(x)}_{|u^+=u^-=u}
=-\fd{W_\rho[\hat u]}{u^-(x)}_{|u^+=u^-=u}.
\ee
The possibility to insert the observable in either time axis is due to the unitarity of the time evolution, the independence of $W_d[\hat u]$ from $t_f$,
\bea
\Tr[U(t'_f,t_i;u)|0\ra\la0|U^\dagger(t'_f,t_i;u)]&=&\Tr[U(t'_f,t_f;u)U(t_f,t_i;u)|0\ra\la0|U^\dagger(t_f,t_i;u)U^\dagger(t'_f,t_f;u)]\nn
&=&\Tr[U(t'_f,t_i;u)|0\ra\la0|U^\dagger(t'_f,t_i;u)].
\eea
We can use this invariance to set $t_f=x^0$ in the definition of the generator function to make the second equation in \eq{eqexpv} obvious. The leading order expression in $u$,
\be\label{kubo}
\la0|\rho(x)|0\ra=-\sum_{\sigma'}\sigma'\int dy D^{\sigma\sigma'}_\rho(x,y)u^{\sigma'}(y),
\ee
where
\be\label{densprop}
D^{\sigma\sigma'}_\rho(x,y)=\fdd{W_d[\hat u]}{u^\sigma(x)}{u^{\sigma'}(y)}_{|\hat u=0},
\ee
cf. Eq. \eq{hamsol} holds for both $\sigma'=+$ and $\sigma'=-$ and Eq. \eq{kubo} is identical with Kubo's linear response formulas \cite{kubo}.

\begin{figure}[ht]
\includegraphics[scale=0.2]{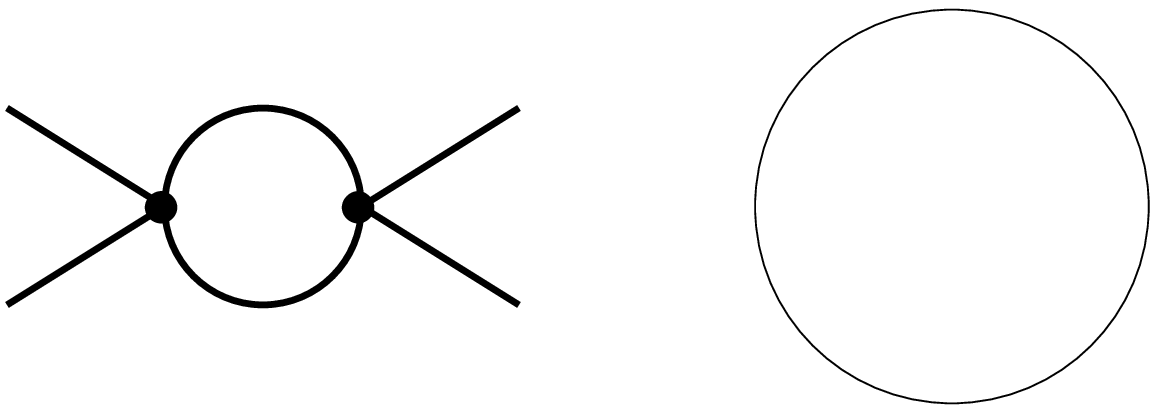}\hskip3cm\includegraphics[scale=0.2]{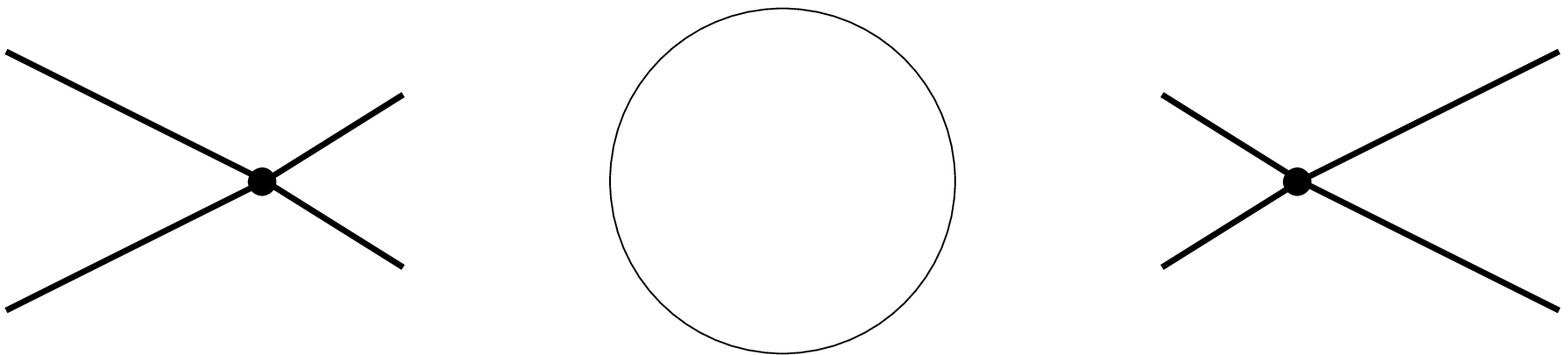}\\(a)\hskip5cm(b)\\
\caption{Leading order graphs, contributing to the expectation value \eq{kubo} with $\sigma'=+$. (a) Factorizable state; (b) entangled state contributions.}\label{linresp}
\end{figure}

It is instructive to identify the physical origin of the causal structure of the propagator \eq{densprop} in perturbation expansion. The leading, $\ord{g_1^2}$ contributions to the expectation value \eq{kubo} with $\sigma'=+$ are given by the graphs of Fig. \ref{linresp}.  The destructive interference between the factorizable state transition amplitude of Fig. \ref{linresp} (a) and the entanglement contribution, depicted in Fig. \ref{linresp} (b), leads to the causal structure in the leading, linear response order. It is easy to see that similar cancellations between virtual excitations (pure state amplitudes) and real excitations (entanglement) are responsible for causality in the higher order of the perturbation series, too.

Unitarity gives the identities
\be
iD_\rho^{++}(x,y)=\la0|T[\rho(x)\rho(y)]|0\ra=\begin{cases}\la0|\rho(x)\rho(y)]|0\ra=iD_\rho^{-+}(x,y)&x^0>y^0\cr
\la0|\rho(y)\rho(x)]|0\ra=iD_\rho^{+-}(x,y)&y^0>x^0\end{cases}
\ee
for the propagator of any local operator, $\rho(x)$, which can be used together with the form $\hD={\cal C}_3[D^n,D^f,D^i]$ to prove directly causality. Causality always follows from a destructive interference between the two time axes for unitary time evolution. Another, stronger result of unitarity, the independence of $W_\rho[u,-u]$ from $t_f$, can be used to prove causality in general, the impossibility that the local potential $u(x)$ influences the expectation value of $\rho(x)$ backward in time: The proof goes by simply setting $t_f=x^0$ in \eq{eqexpv} \cite{ed}.

Despite this general causality argument the CTP formalism lays bare that causality, the cause preceding the effect, is not always automatic. In fact, let us make a perturbation on our classical system at time $t_p$: The trajectory $\tilde x(\tilde t)$, introduced in Sec. \ref{initcs}, is perturbed twice, at $\tilde t_1=t_p$ and $\tilde t_2=2t_f-t_i-t_p$, and it is not obvious that no effect is left for $\tilde t<\tilde t_1$ or $\tilde t>\tilde t_2$, as expected for a causal system. Nevertheless, the simple numerical integration of the Newton or the Schr\"odinger equation proves causality for the initial condition problem of a finite classical system beyond any possible doubt because the values of an external source can influence the solution only after having been used in the integration. How could this argument become invalid for an infinite system? The answer comes from a better defined, regulated setting of this problem, where the numerical integration is performed as a successive, $t\to t+\dt$, solution of a finite difference equation with $\dt>0$. We are faced here with two limits, the continuum limit, $\dt\to0$, and the thermodynamical limit, where the number of degrees of freedom tends to infinity. The dynamics of an infinite system is realized by carrying out the thermodynamical limit first. But the causality is assured by the numerical integration only if the limits are carried out in the opposite order and a possible noncommutativity of the limits opens the way to acausality \cite{effthpch,irrev}, in a manner reminiscent of phase transitions. 

There is still another issue to settle for acausal theories. Since unitarity was used to prove that the dynamics is causal, one suspects nonunitary time evolution in acausal theories. But the unitarity of the time evolution is essential in the CTP formalism to carry out the limit $t_f\to\infty$. In fact, the $t_f$ dependence in the generator functional renders the CTP dynamics trivial for $t_f=\infty$ for nonunitary time evolution. How can a non-trivial quantum theory be acausal? The answer is given by a simple harmonic model \cite{irrev}, with a condensation point in its spectrum. This spectrum requires infinitely long observations to resolve the dynamics of each degree of freedom. The observations, carried out in an arbitrary long but finite amount of time miss infinitely many degrees of freedom and cannot give account of their dynamics, in particular, the unitarity cannot be verified in the manner, mentioned above, for $t_f-t_i<\infty$. It is worthwhile mentioning that irreversibility appears in a similar manner.

\subsection{Quantum-classical transition}\label{qctrs}
We close with a few qualitative remarks about the quantum-classical transition. 

{\em Decoherence:} The first point concerns decoherence, the suppression of the off-diagonal matrix elements of the reduced density matrix, which is a necessary condition of the classical limit \cite{zeh,zurek}. The reduced density matrix, being Hermitian, is always diagonalizable but the decoherence in a given basis is a well-defined, nontrivial problem: It addresses the classical limit of observables that are diagonal in the basis in question. 

A slight generalization of the CTP method, the open time path scheme gives access to the reduced density matrix. This scheme is based on the generator functional,
\be
e^{\ih W[\hj;\phi^+_f,\phi^-_f]}=\la\phi^+_f|\Tr_\psi\left[T\left[e^{-\ih\int dx[H^+(x)-\phi^+(x)j^+(x)]}\right]|0_p\ra\la0_p|T^*\left[e^{\ih\int dx[H^-(x)-\phi^-(x)j^-(x)]}\right]\right]|\phi^-_f\ra,
\ee
where $\phi^\pm(\v{x})$ label the reduced density matrix elements and the trace is taken over the Fock space of the $\psi$ field. In the path integral formula, 
\be\label{reddm}
e^{\ih W[\hj;\hphi_f]}=\int D[\hphi]D[\hpsi]e^{\ih S[\phi^+,\psi^+]-\ih S[\phi^-,\psi^-]+\ih S_{BC}[\hphi]+\ih S_{BC}[\hpsi]+\ih\hphi\hs\hj},
\ee
we integrate over system field configurations that follow an open path, $\hphi(t_f,\v{x})=\hphi_f(\v{x})$, while the environment paths remain closed, $\psi^+(t_f,\v{x})=\psi^-(t_f,\v{x})$. One can define the effective theory for the reduced density matrix, as well, by the help of Eqs. \eq{qinffl} and \eq{inflqft},
\be\label{ereddm}
e^{\ih W[\hj;\hphi_f]}=\int D[\hphi]e^{\ih S_{eff}[\hphi]+\ih\hphi\hs\hj}.
\ee
It is an important simplification that the finite part of the effective action is independent of the matrix elements, $\phi^\pm_f$, and it agrees with the effective action, obtained in the CTP formalism. The only difference between the generator functionals \eq{ctpefgf} and \eq{ereddm} is the final conditions on the field configuration in the integration at $t_f<\infty$ or the presence or absence of $S_f[\hphi]$ in the action for $t_f=\infty$.

The density matrix characterizes the state of the system at a given time hence decoherence refers to a given, instantaneous state, too. The way it emerges dynamically during the time evolution can be traced by inspecting $\Im S_{eff}$, the part of the effective action that controls the magnitude of the density matrix element: The decoherence of the system field, $\phi(x)$, is described by the growth of $\Im S_{eff}[\phi,\phi^d]$ as $|\phi^d|$ is increased \cite{griffith,omnes,gellmann,dowker}. There are two types of contributions in the $\ord{\phi^{d2}}$ part of the effective action \eq{eakp}: The contributions of functional derivatives, acting on $\Im S_2$, describe the environment induced decoherence since $S_2$ represents the system-environment entanglement. The term with the functional derivatives of $\Im S_1$ shows irreversibility induced decoherence because the imaginary part of the action indicates a finite lifetime of the excitations, the leakage of the system into the environment. In the harmonic toy model, \eq{ftoyact}, the structure $\hD={\cal C}_2[D^n,D^f]$ of the propagators makes these two contributions to decoherence equivalent, $\Im K^{++}_{eff}=\Im K^{-+}_{eff}$, underlying the identical dynamical origin of irreversibility ($\Im K^{++}_{eff}$) and decoherence ($\Im K^{-+}_{eff}$). It is important to check the sign of $\Im K^{-+}_{eff}$. The quadratic form of the $\ord{\phi^{d2}}$ part of the effective action \eq{effactoyf} is $-ig^2D^i_e$, ignoring $\ord{\epsilon}$ terms. The inequality \eq{dibound} shows that $\Im S_{eff}[\hphi]$ indeed suppresses, rather than enhances, the contributions of separated CTP pair trajectories in this toy model. 

{\em Entanglement:} The system-environment entanglement in a pure system plus environment state makes the system state mixed. Assuming a pure initial state for the system and the environment the reduced density matrix, given by Eq. \eq{ereddm}, describes a mixed state if and only if $S_2[\hphi]\ne0$. Therefore, the couplings between the two CTP copies, or nonholonomic forces, represent the system-environment entanglement.

We are now in the position to state the strengths of the CTP formalism: Its strategy of constructing open, effective dynamics runs parallel in the classical and quantum domains. In classical mechanics one starts with the system and environment coordinates $(x,y)$, introduced for a closed system with holonomic forces, and the elimination of the environment leaves behind two copies of the system, described by the coordinates $(x^+,x^-)$. A pure system-environment state of quantum mechanics is identified by a wave function $\Psi(x,y)$ and after the elimination of the environment one is left with mixed system states, described by a density matrix $\rho(x^+,x^-)$. The reduplication of the degrees of freedom is rather unusal, because it corresponds to nondefinite kinetic energy in classical physics and the modification of the expression of probability in the quantum case. 

Another advantage of the CTP formalism becomes clear when the perturbation series of the expectation values is visualized by means of Feynman graphs. This scheme reproduces the complexity of the system-environment entanglement with the simplicity of Feynman's view of elementary processes in space-time.

{\em Fluctuations:} The classical and the quantum fluctuations have different origins: the former signals a lack of information and the latter is the manifestation of quantum uncertainties, imposed by the canonical commutation relations. The fluctuations are introduced in the deterministic classical mechanics by some probability distribution of the initial conditions, the identical modification of the initial conditions for the two CTP copies. Thus completely decohered, classical fluctuations appear in $\phi=(\phi^++\phi^-)/2$ and leave $\phi^d=\phi^+-\phi^-$ unchanged. 

The quantum fluctuations at $t_i$ in the generator functional \eq{ctpqmgf}, assumed to contain a factorizable density matrix, are introduced by the independent integration over the initial values of the pair of CTP trajectories. Hence the quantum fluctuations in a pure state are represented by the independent, uncorrelated fluctuations of the two CTP copies. If there is an environment to interact with, then the resulting decoherence gradually correlates the fluctuations of the two CTP copies and in the limit of strong decoherence, $\phi^d\to0$, the fluctuations are indistinguishable from the classical one. Though there is a weak decoherence even in the absence of environment, generated by $K_0^i$ of Eq. \eq{ctpscpr} that represents the closing of the two CTP trajectories at the distant future, such an  infinitesimal effect alone supports no correlations between the CTP copies at finite time.

It is easy to find the origin of classical fluctuations in a strongly decohered system. The quantum initial condition, set by a pure initial state, determines the initial coordinate and velocity on the level of averages only. The fluctuations of the coordinate (momentum) are encoded in the diagonal (off-diagonal) elements of the density matrix, given in coordinate representation and the decoherence of the coordinate influences them in a different manner. When strong decoherence sets $x^d\sim0$, then information about the momentum is lost but the coordinate fluctuations, arising from the initial state, are not suppressed. Thus the classically looking fluctuations of the strongly decohered system originate from the quantum fluctuations in the initial pure state. 

{\em Quantum-classical transition:} The classical limit is usually presented as the dominance of the path integral expression of the matrix elements of the time evolution operator \eq{ctpgfncto} by paths in the vicinity of the classical trajectory, an approximation that can be justified in the limit $\hbar\to0$. This picture suggests that the classical limit is ``rigid'' when the observables receive contributions from the vicinity of the classical trajectory. Though being a correct mathematical limit to approximate integrals, it cannot be the true classical limit, and the latter is unobservable in a single transition amplitude between pure states, without taking into account decoherence, which is a necessary condition of the classical limit.

One expects to recover classical physics from quantum mechanics when the energy levels are very close to each other and the system averages itself quickly over a large number of stationary states during the time evolution. Since the stationary states are orthogonal to each other, the system needs very small energy to orthogonalize itself in this limit, which therefore should rather be qualified as ``soft.''

The strong decoherence limit of the CTP path integrals is nontrivial, the action \eq{ctpactspl} is prevented from being vanishing identically for $x^d=0$ by the infinitesimal splitting term, $S_{spl}[\hx]$, only. Such an almost degeneracy, the choice of the minus sign in the right hand side of Eq. \eq{ctps}, is crucial in establishing the desired correlation between the CTP copies, as was pointed out in Sec. \ref{cgreens}. Therefore the approach of classical physics cannot be followed in a simple, formal manner.

Nevertheless, one can make a small, qualitative step toward the strong decoherence limit for weak system-environment coupling. The $\phi$ dependent part of the action \eq{effackp} is $\ord{\phi^d}$ ($\Re S_1$), and $\ord{g}$ ($\Im S_1$, $S_2$), where $g$ stands for a generic system-environment coupling constant. Therefore the $\phi$ dependence of the integrand of the path integral is weak. Hence the ``restoring force to equilibrium'' is weak during the time evolution and the fluctuations of $\phi(x)$ are large for a strongly decohered system, in other words the classical limit is indeed soft.

\section{Summary}\label{sums}
The well known CTP formalism is extended to classical mechanics in this work. A unified description of an open system, considered as a subset of a closed dynamical system, is discussed in the classical and quantum domains where the nonholonomic effective classical forces and the system-environment entanglement of the quantum state are handled by a reduplication of the degrees of freedom. The two copies (i) are placed into a symplectic structure or into complex conjugate representations in the classical and quantum cases, respectively, (ii) obey the same initial conditions, and (iii) are constrained to assume the same position, to be determined by the dynamics, at the final time. This scheme reveals a rather surprising possibility, the mapping of the system-environment interactions into the interactions of two copies of the system. One would have thought that the system-environment interactions reflect the richness of the usually large environment. The fact that this is not the case and the complexity of the system-environment interactions is limited by the system alone might be accepted by noting that the reaction of a simple system to a complex environment should remain simple. 

The reduplication of the degrees of freedom allows us to establish an action principle for initial condition problems and for dissipative forces in classical mechanics and to preserve the path integral formalism and the intuitive appeal of Feynman graphs in representing the perturbation series of expectation values for an open quantum system. It is shown that this scheme goes beyond the traditional effective theories, used in quantum field theory, by including processes that leave the environment in an excited state. 

The distinguishing feature of this scheme, without analogy in the traditional action formalism, is the interaction between the CTP copies. It stands for the unrestricted, open ended time evolution of the environment within the action principle and makes the dynamics open. In case of the effective theory of charges in classical electrodynamics the coupling between the CTP copies is the interaction of the charges by the far radiation field. The very same coupling between the copies represents the system-environment entanglement in the quantum case. Since the entanglement contribution to observables is $\ord{\hbar^0}$, it is natural to find a remnant of entanglement in the classical domain.

Effective quantum field theories are widely used from condensed matter physics to high energy physics because we have no realistic hope to discover and test experimentally fundamental, elementary theories. The extension of the current technique of effective theories by means of the CTP formalism is necessary to cover diffusion, irreversibility, acausality, decoherence and other phenomena that rely on soft collective excitations of the environment. We repeat that the quantum CTP formalism has already been well established, and the only new elements of this work are its relation to classical mechanics and the necessity of its application in effective theories.

The transmutation of quantum fluctuations into a classical one can be followed qualitatively in strongly decohered systems. The quantum fluctuations in a pure initial state appear as independent fluctuations of the CTP copies. If strong decoherence builds up during the time evolution then the fluctuations of the CTP copies become identical and indistinguishable from classical fluctuations. This scenario is in agreement with the well known peculiarity of the density matrix, namely that the very same density matrix can be obtained in two different manners: On the one hand, to reproduce the expectation values of a system that is entangled with another one, we have to use (reduced) density matrices. On the other hand, incomplete knowledge of the quantum state can be taken into account by using density matrices. The transfer of the unsuppressed part of the quantum fluctuations of a strongly decohered system into the form of classical fluctuations underlines indeed such a dual role of classical uncertainties.

The trajectories of strongly decohered systems occupy a rather singular region in the space of trajectories: The trajectories of the two CTP copies are almost identical and the dependence of the action on the remaining common trajectory is weak. This situation is reminiscent of the strong coupling regime of quantum field theory and may render the comparison of the classical and quantum domains highly nontrivial. For instance, the correspondence principle is based on the assumption that the degrees of freedom are identical in the quantum and the classical regimes. The phenomenon of quark confinement demonstrates that a strong coupling regime may separate regimes, governed by significantly different degrees of freedom. A nonperturbative treatment of the CTP formalism is needed to derive classical physics from quantum theory and to establish the correspondence principle in a systematic manner.

Beyond these qualitative remarks one expects other issues, as well, where a CTP effective theory might be a useful tool. The systematic treatment of open systems, blended with the renormalization group method should lead us to a theory of open critical systems and an extension of the traditional classification of operators around the critical points. Furthermore, the resolution dependence of an effective CTP theory should expose a new crossover in any system without a gap in its excitation spectrum, where the quantum physics turns into a classical one. A family of problems might be addressed in this manner, for instance, a more realistic treatment of particle detection in a cloud or wire chamber in the laboratory, or on a cosmological scale, the formation and decoherence of density and gravitational fluctuations in inflation. The CTP formalism of general gelativity contains two space-times and should place the analytic extension of the Schwarzschild metrics, discovered in the Kruskal-Szekeres coordinate system, in a wider context. Finally, the more efficient treatement of time reversal odd interactions may lead us to a better understanding of the origin of the CP-violating sector of the Standard Model and its the impact on low energy and classical phenomena.

\appendix
\section{Noether theorem}\label{nothera}
The conservation of momentum and energy is discussed here briefly in case of the semiholonomic forces are acting. The definition of infinitesimal symmetry transformation is the same in CTP as in the usual case: namely the action is allowed to be changed at most by a total time derivative, a boundary term. The effective action is usually nonlocal in time, but a useful approximation scheme is the gradient expansion where one assumes the existence of a local effective Lagrangian with possible higher order derivatives: the simple form $L=L_1(x^+,\dot x^+)-L_1(x^-,\dot x^-)+L_2(\hx,\dot{\hx})$ will be used below.

To check the momentum conservation we perform the infinitesimal translation $\hx\to\hx+\epsilon\hat a$ with time independent $\epsilon$ and $\hat a$. This transformation applies to the system only, and the translation invariance of the effective dynamics may be broken by the environment initial conditions. To test such a symmetry breaking $\epsilon$ is made time dependent and its Lagrangian,
\bea
L(\epsilon,\dot\epsilon)&=&L(\hx+\epsilon\hat a,\dot{\hx}+\dot\epsilon\hat a)\nn
&=&\epsilon\hat a\fd{L}{\hat x}+\dot\epsilon\hat a\fd{L}{\dot{\hat x}}+\ord{\epsilon^2},
\eea
can be used to arrive at the equation of motion,
\be\label{epseom}
0=\hat a\left(\fd{L}{\hat x}-\frac{d}{dt}\fd{L}{\dot{\hat x}}\right),
\ee
which is satisfied if $\hx(t)$ is a solution of the equation of motion of the effective Lagrangian. The holonomic forces are included in $L_1$, and thus it is natural to define the momentum by the equation
\be
P=\sum_\sigma\sigma a^\sigma\fd{L_1(x^\sigma,\dot x^\sigma)}{\dot x^\sigma}
\ee
and Eq. \eq{epseom} leads to the balance equation,
\be\label{pnonc}
\dot P=\hat a\left(\fd{L}{\hat x}-\frac{d}{dt}\fd{L_2}{\dot{\hat x}}\right).
\ee

The simplest choice is $a^+=a^-=1$, but the corresponding momenta are vanishing for the solution of the equation of motion. It is more useful to perform the translation on one copy of the system only,  and use $a^+=1$, $a^-=0$, which does not correspond to a symmetry transformation, but
\be
P=\fd{L_1}{\dot x}
\ee
is the usual momentum. The rate of its change is
\be
\dot P=\fd{L_1}{x}-\frac{d}{dt}\fd{L_2}{\dot x^+}_{|x^+=x^-=x},
\ee
according to Eq. \eq{pnonc}. The first term in the right hand side stands for the lack of translation invariance of the holonomic forces and the remaining terms represent the semiholonomic forces which make the momentum time dependent and system dynamics open. In the case of the Lagrangian \eq{newtonfr} the balance equation for momentum, $P=m\dot x$, is $\dot P=-k\dot x-U'(x)$.

To test energy conservation we make a variation of the trajectory $x^+(t)$ which is induced by a time dependent translation in time, $\delta x^+(t)=-\epsilon\dot x(t)$, $\delta x^-(t)=0$ and find the $\ord{\epsilon}$ part of the Lagrangian for $\epsilon$,
\be
L(\epsilon,\dot\epsilon)=-\epsilon\dot x^+\fd{L}{x^+}-\epsilon\ddot x^+\fd{L}{\dot x^+}-\epsilon\frac{\partial L}{\partial t}-\dot\epsilon\dot x^+\fd{L}{\dot x^+}+\epsilon\frac{\partial L}{\partial t},
\ee
where the time dependence, detected by $\partial/\partial t$, comes through $x^-(t)$. The sum of the first three terms gives a total derivative,
\be\label{ecepsl}
L(\epsilon,\dot\epsilon)=-\epsilon\left[\frac{dL}{dt}-\frac{d}{dt}\left(\fd{L}{\dot x^+}\dot x^+\right)\right]-\frac{d}{dt}\left(\fd{L}{\dot x^+}\epsilon\dot x^+\right)+\epsilon\frac{\partial L}{\partial t}.
\ee
The rate of change of the energy,
\be
H=\frac{\partial L_1}{\partial\dot x}\dot x-L_1,
\ee
can be written in the form
\bea
\frac{d}{dt}H(x^+,\dot x^+)&=&-\frac{\partial L}{\partial t}+\frac{d}{dt}\left(L_2-\fd{L_2}{\dot x^+}\dot x^+\right)\nn
&=&\left(\fd{L_2}{x^+}-\fdd{L_2}{\dot x^+}{x^+}\dot x^+-\fdd{L_2}{\dot x^+}{\dot x^+}\ddot x^+\right)\dot x^+
\eea
with the help of the equation of motion of the Lagrangian \eq{ecepsl}. In case of the Lagrangian $L_1=m\dot x^2/2$, $L_2=g(x^-)\dot x^+-g(x^+)\dot x^-$, where the stability of the equilibrium position requires $g'(x)\ge0$, one finds
\be
\frac{d}{dt}H=-g'(x)\dot x^2\le0.
\ee
The nontrivial point here is that the Noether theorem can be used to describe energy nonconservation even though the Lagrangian is invariant under translation in time.

\section{Relative state}\label{relsta}
The mixed nature of the system state is the easiest to follow with the help of the relative state, introduced by Everett, \cite{everett}. The space of states of our full system is a direct product, ${\cal H}={\cal H}_s\otimes{\cal H}_e$, where ${\cal H}_s$ and ${\cal H}_e$ denote the system and environment state spaces and we shall use the basis sets $|p\ra\in{\cal H}_s$, $|q\ra\in{\cal H}_e$ and $|p\ra\otimes|q\ra=|p,q\ra\in{\cal H}$. We assume furthermore that the system is in a pure state, $|\Psi\ra$, and define its relative state, corresponding to an environment state $|\psi\ra\in{\cal H}_s$ as
\be
|R(\chi)\ra=N(\chi)\sum_p|p\ra\la p|\otimes\la\chi|\Psi\ra,
\ee
where $N(\chi)>0$ is chosen for normalization. It is easy to see that this definition is unique, i.e. it is actually independent of the choice of the system basis.

The relative state encodes the conditional system expectation values. In fact, let us write the full system state by using our basis as
\be
|\Psi\ra=\sum_{pq}|p,q\ra\la p,q|\Psi\ra
\ee
and introduce the relative state for each environment basis element,
\be
|R(q)\ra=N(q)\sum_p|p\ra\la p,q|\Psi\ra
\ee
where $N(q)$ is given by the equation
\be
\frac1{N^2(q)}=\sum_p|\la p,q|\Psi\ra|^2=P(q),
\ee
$P(q)$ being the probability of finding the environment basis vector $q$ in our state. The decomposition of the full system state,
\be
|\Psi\ra=\sum_q\sqrt{P(q)}|R(q),q\ra
\ee
indicates that each environment state forms its own counterpart, a relative system state. A more detailed picture is found by defining the conditional probability, $P(p|q)=|\la p,q|\Psi\ra|^2/P(q)$, and considering a system observable that is diagonal in the basis, $A|p\ra=\lambda(p)|p\ra$. The expectation values in a relative state,
\be
\la R(q)|A|R(q)\ra=\sum_p\lambda_pP(p|q),
\ee
and in the full system state,
\be\label{fsev}
\la\Psi|A|\Psi\ra=\sum_qP(q)\la R(q)|A|R(q)\ra,
\ee
confirm the interpretation of relative states as conditional pure system states.

Since the system-environment interactions make the system, depending on the environment there should be several linearly independent relative system states and system-environment entanglement appears. There are no conditional states anymore, and the expectation value \eq{fsev} cannot be reproduced for an arbitrary observable $A$ by any fixed system state vector in that case. In fact, let us suppose the contrary, that there exists a state vector, $|\phi\ra$, and consider the measurement of the projection operator, $A=|\phi\ra\la\phi|$. The result must be $1$ but Eq. \eq{fsev} gives less than $1$ unless there is a single relative state only. Hence one needs density matrix
\be\label{densm}
\rho=\sum_q|R(q)\ra P(q)\la R(q)|
\ee
to represent the averages, $\la A\ra=\Tr[A\rho]$, for an interacting system and environment. 

Note that the system-environment entanglement is encoded in the spread of the distribution of the environment quantum numbers. Hence, symmetries and the following selection rules may restrict seriously the amount of entanglement. For instance, if the sum $p+q=P$ is conserved then there is a single relative state only.

\end{document}